%% file: distance_revise2.tex
\documentclass[twocolumn]{aastex6}

\begin{document}
\title{THE GOULD'S BELT DISTANCES SURVEY (GOBELINS) II. \\Distances and structure towards the Orion Molecular Clouds}

\author{Marina Kounkel\altaffilmark{1}, Lee Hartmann\altaffilmark{1}, Laurent Loinard\altaffilmark{2,3}, Gisela N. Ortiz-Le\'{o}n\altaffilmark{2}, Amy J. Mioduszewski\altaffilmark{4}, Luis F. Rodr\'{i}guez\altaffilmark{2,5}, Sergio A. Dzib\altaffilmark{3}, Rosa M. Torres\altaffilmark{6}, Gerardo Pech\altaffilmark{2}, Phillip A. B. Galli\altaffilmark{7,8,9}, Juana L. Rivera\altaffilmark{2}, Andrew F. Boden\altaffilmark{10}, Neal J. Evans II\altaffilmark{11}, Cesar Brice\~{n}o\altaffilmark{12}, John J. Tobin\altaffilmark{13,14}}
\altaffiltext{1}{Department of Astronomy, University of Michigan, 1085 S. University st., Ann Arbor, MI
48109, USA}
\altaffiltext{2}{Instituto de Radioastronom\'{i}a y Astrof\'{i}sica, Universidad Nacional Aut\'{o}noma de Mexico, Morelia 58089, Mexico}
\altaffiltext{3}{Max Planck Institut f\"{u}̈r Radioastronomie, Auf dem H\"{u}̈gel 69,D-53121 Bonn, Germany}
\altaffiltext{4}{National Radio Astronomy Observatory, Domenici Science Operations Center, 1003 Lopezville Road, Socorro, NM 87801, USA}
\altaffiltext{5}{King Abdulaziz University, P.O. Box 80203, Jeddah 21589, Saudi Arabia}
\altaffiltext{6}{Centro Universitario de Tonal\'{a}, Universidad de Guadalajara, Avenida Nuevo Periférico No. 555, Ejido San Jos\'{e}, Tatepozco, C.P. 48525, Tonal\'{a}, Jalisco, M\'{e}xico}
\altaffiltext{7}{Univ. Grenoble Alpes, IPAG, 38000, Grenoble, France}
\altaffiltext{8}{CNRS, IPAG, F-38000 Grenoble, France}
\altaffiltext{9}{Instituto de Astronomia, Geof\'isica e Ci\^encias Atmosf\'ericas, Universidade de S\~ao Paulo, Rua do Mat\~ao, 1226, Cidade Universit\'aria, 05508-900, S\~ao Paulo - SP, Brazil}
\altaffiltext{10}{Division of Physics, Math and Astronomy, California Institute of Technology, 1200 East California Boulevard, Pasadena, CA 91125, USA}
\altaffiltext{11}{Department of Astronomy, The University of Texas at Austin, 2515 Speedway, Stop C1400, Austin, TX 78712-1205, USA}
\altaffiltext{12}{Cerro Tololo Interamerican Observatory, Casilla 603, La Serena, Chile}
\altaffiltext{14}{Homer L. Dodge Department of Physics and Astronomy, University of Oklahoma, 440 W. Brooks Street, Norman, OK 73019, USA}
\altaffiltext{14}{Leiden Observatory, Leiden University, P.O. Box 9513, 2300-RA Leiden, The Netherlands}
\email{mkounkel@umich.edu}

\begin{abstract}
We present the results of the Gould's Belt Distances Survey (GOBELINS) of young star forming regions towards the Orion Molecular Cloud Complex. We detected 36 YSOs with the Very Large Baseline Array (VLBA), 27 of which have been observed in at least 3 epochs over the course of 2 years. At least half of these YSOs belong to multiple systems. We obtained parallax and proper motions towards these stars to study the structure and kinematics of the Complex. We measured a distance of 388$\pm$5 pc towards the Orion Nebula Cluster, 428$\pm$10 pc towards the southern portion L1641, 388$\pm$10 pc towards NGC 2068, and roughly $\sim$420 pc towards NGC 2024. Finally, we observed a strong degree of plasma radio scattering towards $\lambda$ Ori.\end{abstract}

\section{Introduction}

Young star forming regions towards Orion have been the subject of much interest, as they are the closest regions of massive young stellar population. The star formation in the Orion Complex is concentrated in two molecular clouds, Orion A and B, with clusters such as the Orion Nebula Cluster (ONC) and L1641 in Orion A, and NGC 2023/2024, NGC 2068/2072 and L1622 in Orion B. These clusters represent the most recent episodes of star formation in the region, which belong to the Orion OB1c and 1d sub-association, containing stars spanning ages from $\sim$1 Myr up to 6 Myr \citep{2008bally}. In addition to the clusters in the main cloud, there are other stellar groups in Orion that host very young stars, like $\sigma$ Ori, in the OB1b sub-association, and the groups of the $\lambda$ Ori association at the northernmost end of the complex. Finally, a somewhat older (8--12 Myr) population is contained within the OB1a sub-association, where most of the parental gas has already been removed.

Over the course of the last century, many attempts have been made to measure distances to the Complex, particularly towards the ONC. Some of the earliest measurements were as high as 2000 pc \citep{1917Pickering} and as low as 185 pc \citep{1918Kapteyn}. Eventually most measurements settled in the 350 to 500 pc range obtained through various means, most typically through zero-age main sequence fitting. Much of the scatter originated from inconsistent assumptions, models, and sample selection \citep[see review by][]{2008Muench}.

For some time, the most widely used distance was 480$\pm$80 pc, obtained from proper motions of H$_2$O masers towards the Orion BN/KL region \citep{1981Genzel}. In the last decade, however, direct stellar parallax measurements of non-thermal emitting masers and stars were made possible through radio Very Long Baseline Interferometry (VLBI). \citet[hereafter MR]{2007menten} obtained a distance of 414$\pm$7 pc from observations of 4 stars - GMR A, F, G, and 12 - in the central (Trapezium) region of the ONC. \citet[hereafter S07]{2007sandstrom} also observed GMR A and obtained a somewhat closer distance of 389$^{+24}_{-21}$ pc. \citet{2007hirota} and \citet{2008kim} observed H$_2$O and SiO masers to obtain a distance of 437$\pm$19 pc and 418$\pm$6 pc respectively in the Orion BN/KL region.

Other major efforts to measure a distance towards the ONC include \citeauthor{2007Jeffries}'s (\citeyear{2007Jeffries}). He used stellar rotation to estimate distances of 440$\pm$34 pc for his entire sample and 392$\pm$32 pc including only stars without active accretion. \citet{2004Stassun} obtained a distance of 419$\pm$21 pc through monitoring the kinematics of a double-line eclipsing binary system, assuming a value for the solar bolometric luminosity of $M_{bol,\odot}=$4.59, although their distance estimate decreased to 390$\pm$21 pc with $M_{bol,\odot}=$4.75. \citet{2009Kraus} obtained a dynamical distance of 410$\pm$20 pc based on modeling the orbit of the close binary $\theta^1$ Ori C. Some attempts have also been made to obtain distances from dust extinction maps towards not just the ONC, but towards several distinct regions in the Orion Complex. \citet{2011Lombardi} estimated 371$\pm$10 pc towards Orion A and 398$\pm$12 pc towards Orion B using extinction maps measured from 2MASS. \citet{2014Schlafly} provided distance estimates of 20 distinct regions through extinction from PanSTARRS photometry, although many of them are highly uncertain.

While the distance measured by MR is currently considered as canonical, it is is based on a small sample of 4 stars. In addition, the MR stars all lie within the central regions of the ONC; however, the Complex spans 100 pc projected on the sky, so it would not be surprising if the different regions of the cloud have substantially different distances, and it would not be surprising if the regions have differing radial distances of the same order. Therefore, even if the distance towards the ONC is known with high accuracy, by applying this distance to other regions an inherent uncertainty of $\sim$20\%, for example, could be introduced, as the Complex is located at the distance of $\sim$400 pc. This propagates to an error of $\sim$40\% in luminosity, to $\sim$70\% in ages of young stars \citep{2001hartmann}.

Currently an ongoing mission of the \textit{Gaia} space telescope is in process of obtaining astrometry towards optically visible sources across the entire sky in order to measure parallaxes accurate to 100 $\mu$as for G$<$17 mag stars \citep{2014deBruijne}, which should provide accuracy in distance measurements to within 5-10\% up to 1 kpc. VLBI observations can provide an important independent check on optical parallax measurements, as shown by the comparison of VLBI with \textit{Hipparcos} distances for the Pleiades \citep{2014Melis}. In addition, radio VLBI can be useful for measuring sources in regions of high extinction and/or significant nebulosity, as is the case in many regions of Orion.

In this paper we present radio VLBI observations of stellar parallaxes of Young Stellar Objects (YSOs) identified towards the Orion Complex, hereby significantly expanding the number of stars in Orion with known distances and kinematics. This work is done as part of Gould's Belt Distances Survey \citep[GOBELINS,][]{gbds}, which is dedicated to measure stellar parallax towards the Ophiuchus \citep{oph2}, Serpens \citep{ser}, Taurus, Perseus, and Orion star forming regions.

\section{Observations}

The observations presented in this paper were made with the National Radio Astronomy Observatory's Very Long Baseline Array (VLBA) at 5 GHz with a 256 MHz bandwidth (spanning the range of 4.852---5.076 GHz). They span a period of 2 years from March 2014 to March 2016, with observations preferentially scheduled near the equinoxes, to target the maxima of the parallactic eclipse along right ascension. All the fields were observed in groups of three per observing session, for a total of 56 observing sessions under the code of BL175 (Table \ref{tab:obs}). Each session was planned as follows: a primary calibrator was alternated between observing each field, and after five iterations encompassing all the fields, three secondary calibrators were observed. The duration of each pointing was $\sim$2 minutes for targeted fields and $\sim$1 minute for each of the calibrators. The geodetic block was observed at the beginning and the end of each session (the frequency of the observations of this block spanned the 4.596---5.076 GHz range). The total observing time was $\sim$1 hour per field. Additionally, we spent 1.5 hour per session on the primary calibrator and 0.1 hour on each of the secondary calibrators.

During correlation each field was reduced to a series of small patches only a few arcseconds in diameter, each patch centered at a phase center corresponding to the targets within a field \citep[a description of the process is presented in][]{oph2}. Targets were identified from the VLA survey of the Orion Complex by \citet{2014kounkel}. In that survey 374 sources were detected, out of which 148 were associated with known YSOs, and 86 were identified as YSO candidates on the basis of their radio properties out of the remaining sources that could not be classified by other means. The criteria for candidacy included a) detectable circular polarization or b) short-term radio variability $>$50\% at either 4.5 or 7.5 GHz.

For the first two epochs of the VLBA Orion observations, 40 fields were observed (Table \ref{tab:obs}). The total number of fields was chosen on the basis of the number of hours awarded for the Gould's Belt Distance Survey. We accommodated observations of all five star-forming clouds targeted by the program, and the field centers were distributed in a way to maximize the number of known YSOs observed. After two epochs we were already able to begin to distinguish between galactic and extragalactic sources on the basis of motion of the sources between epochs. Twelve fields where no galactic sources have been detected were removed from the survey, six new fields were added to include more isolated YSOs from the VLA survey. The number of fields has been further cut to only 26 for epoch 4, and to 17 for epoch 5. 

The data were reduced in AIPS \citep{aips} following the standard prescription for the VLBA data. The multi-band delays were removed using the DELZN task based on the geodetic sources \citep{2004reid}. The phase gradient across the sky was then calibrated using secondary calibrators with the ATMCA task. When multiple sources were observed in the same field, the same calibration was applied to all sources. Positions of all the sources were referenced to the primary calibrator. Finally, all sources were imaged and the positions of all point sources were extracted using task the JMFIT task. More details on data reduction are presented in \citet{oph2}.

\begin{figure}
\plotone{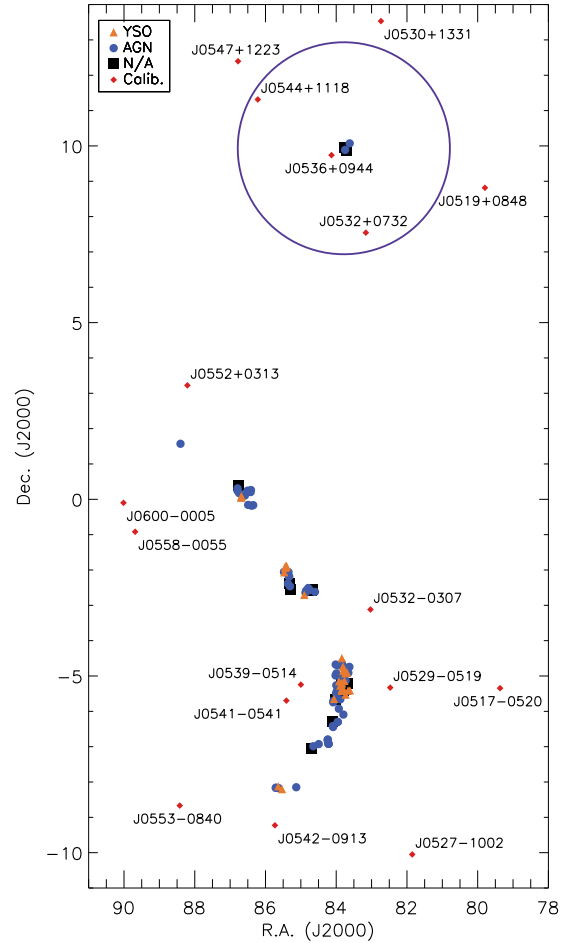}
\caption{Locations of the sources observed by this program. They are separated into the confirmed YSOs, sources that are most likely associated with the AGN activity, and the ones that could not be identified as either due to either insufficient number of epochs observed, or due to poor astrometry. Locations of all the calibrators are also included with their names. The circle at the top of the figure is 3$^\circ$ in radius, and it shows the approximate position of the $\lambda$ Ori ring. \label{fig:all}}
\end{figure}

Positions of all sources were referenced relative to the primary calibrators (Table \ref{tab:poscalib}). The arrangement of the primary and secondary calibrators for each field is described in Table \ref{tab:calib}. This configuration was preserved through all the epochs. The exception to this are the $\lambda$ Ori fields, as their primary and secondary calibrators were extended, which resulted in uncertain astrometry. We propose that plasma scattering from the supernova bubble 2.5--3$^\circ$ in radius around $\lambda$ Ori is responsible for the image blurring (See Appendix A). An attempt was made to switch to a different nearby calibrator that would improve the astrometry. However, as the assumed absolute positions of the calibrators are not referenced relative to each other, a positional offset was introduced to the sources in the field.

A few other calibrators do have some structure, most likely due to jet activity, however, an evolving jet structure should not significantly influence measured parallaxes \citep{2007menten}, particularly if the main source is point-like and the jet emission is sufficiently displaced. The two most notable calibrators with structure are J0539-0514 and J0532-0307. The former produced many errors in the calibration solutions as it was not bright enough during the first two epochs, with a typical flux $\sim$30 mJy and substructure was not immediately apparent, but during the third epoch its flux had increased to $\sim$50 mJy, resulting in a significant improvement in the calibration, and there was a clear emission from a second component at position angle $\sim$240$^\circ$, $\sim$1 mas away; this emission persisted during all the remaining epochs. The latter calibrator, J0532-0307, always had a spatially resolved second component at p.a. $\sim$150$^\circ$, $\sim$10 mas away.

Our VLBA survey fields covered a total of 300 sources from the VLA survey. As VLBA detections generally require non-thermal emission due to their high brightness temperatures, only 116 objects have been detected (although some of them can be resolved into multiple objects or a jetted structure). We report only on objects that either a) were detected in at least two epochs or b) had a single detection $>5\sigma$ (Figure \ref{fig:all}). Most likely the remaining undetected objects emit only thermally. Thirty six of the detected systems can be definitely identified as YSOs on the basis of astrometric motion, and 57 are most likely associated with AGN activity. The remaining 23 objects have detection in only one epoch or have astrometry too poor to make a definitive determination (Table \ref{tab:pos}).

Out of 148 objects in the VLA survey that were known YSOs in the literature, 36 have been detected with VLBA. Three of these sources have been falsely identified, as we cannot confirm their membership to the Orion Complex on the basis of their astrometry. GBS-VLA J054121.69-021108.3 (=VLBA 55), GBS-VLA J053542.27-051559.3 (=VLBA 110), and	GBS-VLA J053532.03-053938.6 (=VLBA 139), all previously identified as YSOs on basis of optical and IR emission, but they do not show a significant astrometric offset between epochs.

\begin{figure}
\plottwo{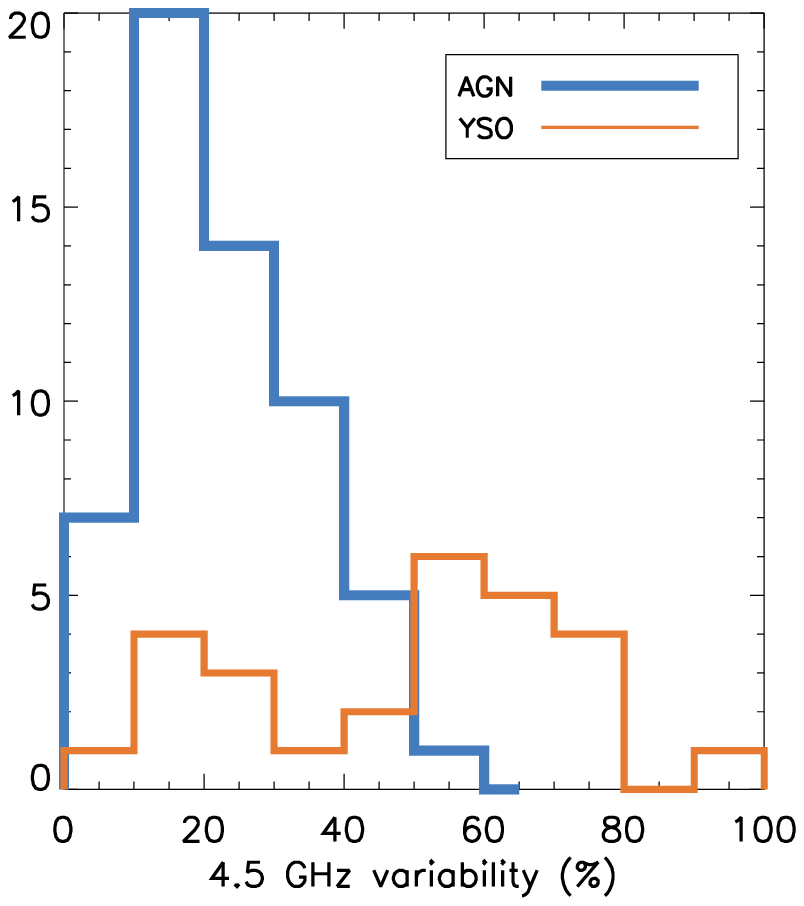}{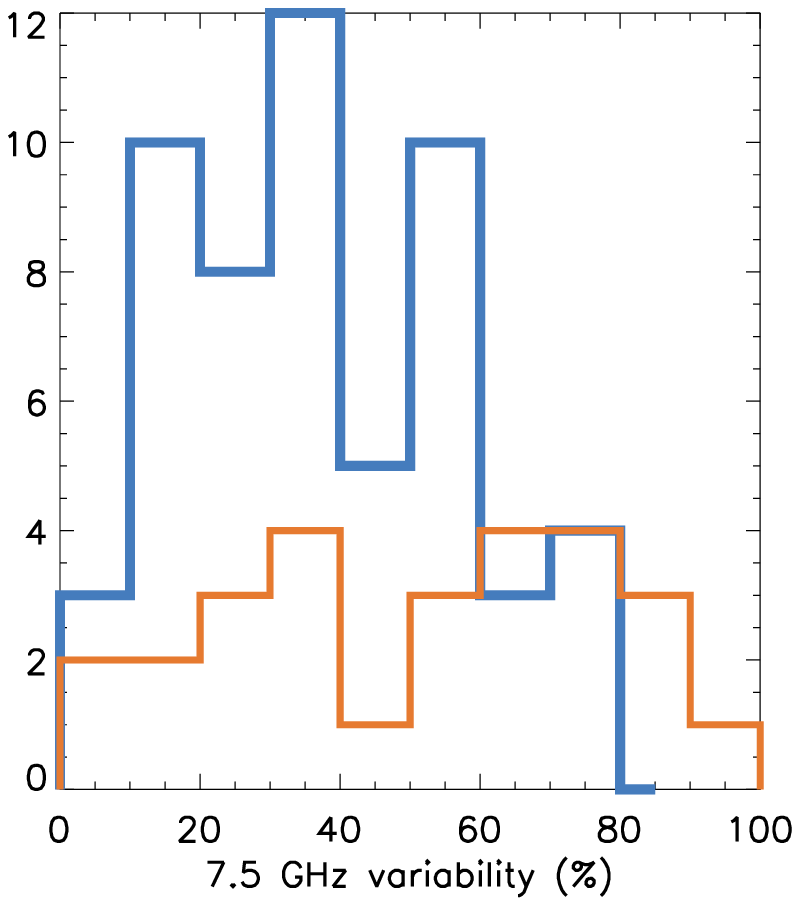}
\caption{Variability reported by \citet{2014kounkel} for all sources that can be distinguished as galactic and extragalactic on the basis of VLBA astrometry.\label{fig:vla}}
\end{figure}

\citet{2014kounkel} identified 86 VLA sources as candidate YSOs, based on their radio properties, and we detected 26 of them with the VLBA.  Only three of these can be confirmed as YSOs in the Orion region. The criteria for selection used by \citet{2014kounkel} appear to be not entirely reliable: while strong variability at 4.5 GHz can indeed be used to distinguish galactic from extragalactic sources, the same cannot be said for 7.5 GHz (Figure \ref{fig:vla}). Similarly, the degree by which circular polarization can be affected by beam squint has been significantly underestimated. Out of nine sources with observed VLA circular polarization detected in this program, we can confirm only two as YSOs.

\section{Fitting}

To fit the parallax and the proper motions, the IDL routine MPFIT \citep{mpfit} was used. This routine fits a given model to a data by minimizing least-squares fit. At the end of each iteration, it outputs only a single array with the weighted differences between the data and the model, and any number of equations can be solved simultaneously.

For a single object, the motion of a star in the plane of the sky is prescribed by
\[\alpha(t)=\alpha_0+\mu_\alpha \cos\delta t+\pi f_\alpha(\alpha,\delta,t)\]
\[\delta(t)=\delta_0+\mu_\delta t+\pi f_\delta(\alpha,\delta,t)\]
\noindent where $\alpha_0$ and $\delta_0$ are positions of the star at a given reference time, and $\mu_\alpha$ and $\mu_\delta$ are the components of the proper motion. $f_\alpha$ and $f_\delta$ are the projections over $\alpha$ and $\delta$ of the parallactic eclipse, and they are given by \citep[e.g.][]{1992Seidelmann}
 \[f_\alpha=(X\sin\alpha-Y\cos\alpha)/\cos\delta\]
 \[f_\delta=X\cos\alpha\sin\delta+Y\sin\alpha\sin\delta-Z\cos\delta\]
\noindent where $X$, $Y$ and $Z$ are the barycentric coordinates of the Earth in units of AU, tabulated using the Python package Skyfield\footnote{\url{http://rhodesmill.org/skyfield/}}.

The uncertainties in the fitted parameters are twofold. First of all, they depend on the positional uncertainties of all the individual detections of the stars as measured by JMFIT, driven by the resolution of VLBA and the flux of the object. This does not take into account possible various systematic offsets in positions between different epochs, which could be significantly larger than nominally quoted positional uncertainties. Typically, the estimation of errors due to systematic offsets is usually done through examining the goodness of the parallactic fit and scaling positional uncertainties until the reduced $\chi^2$ of the fit becomes equal to 1 \citep[e.g.][]{2007sandstrom,2007menten}.

Approximately half of GK main sequence stars and 30\% of M stars belong to a multiple systems \citep{1991Duquennoy,1992Fischer,2010Raghavan,2013duchene}, and the motion of the binary projected onto the plane of the sky can degrade the goodness of the fit. These multiple systems can be roughly divided into three categories, based on the effect they have on parallax and proper motion fit. 

\begin{enumerate}

\item Binaries with orbital periods much longer than the total monitoring time covered by this program (e.g. $>>$10 years). As the star would only barely move in its orbit, this motion would be approximately linear. It is possible to introduce and fit for an acceleration term to correct for the minor shifts due to non-linearity. Determination of the parallax should not be affected by these binaries. Proper motion would not represent the true proper motion of the system, as it is strongly affected by the orbital motion of the star.

\item Binaries with intermediate orbital periods. The effects of the binary motion cannot be ignored during the parallactic fit due to the noticeably changing acceleration of the star; therefore, it is necessary to fit the Keplerian parameters for the binary and the parallax simultaneously. The main orbital parameters are the semi-major axis $a_1$ of the primary, the orbital period $P$, eccentricity $e$, argument of the pericenter $\omega$, the time of passage of the pericenter $T_P$, inclination $i$, argument of the ascending node $\Omega$ and, in case of the astrometric binary with both components detected, mass ratio $q$. The mean anomalies for the dates of observations are calculated with a given $P$ and $T_P$. Then a true anomaly $\theta$ and a radius from the center of mass $r$ are determined along a Keplerian orbit with a given \textit{e} for the positions corresponding to these mean anomalies. This orbit is scaled and projected onto the plane of the sky through
\[\alpha(t)=a_1r(\cos(\theta+\omega)\sin\Omega-\sin(\theta+\omega)\cos\Omega\cos i)/\cos\delta\]
\[\delta(t)=a_1r(\sin(\theta+\omega)\sin\Omega\cos i + \cos(\theta+\omega)\cos\Omega)\]

For secondary stars, $\theta$ is rotated by 180$^\circ$, and $a_2$ is used instead, which is scaled from $a_1$ by $q$.

\begin{figure}
	\gridline{\fig{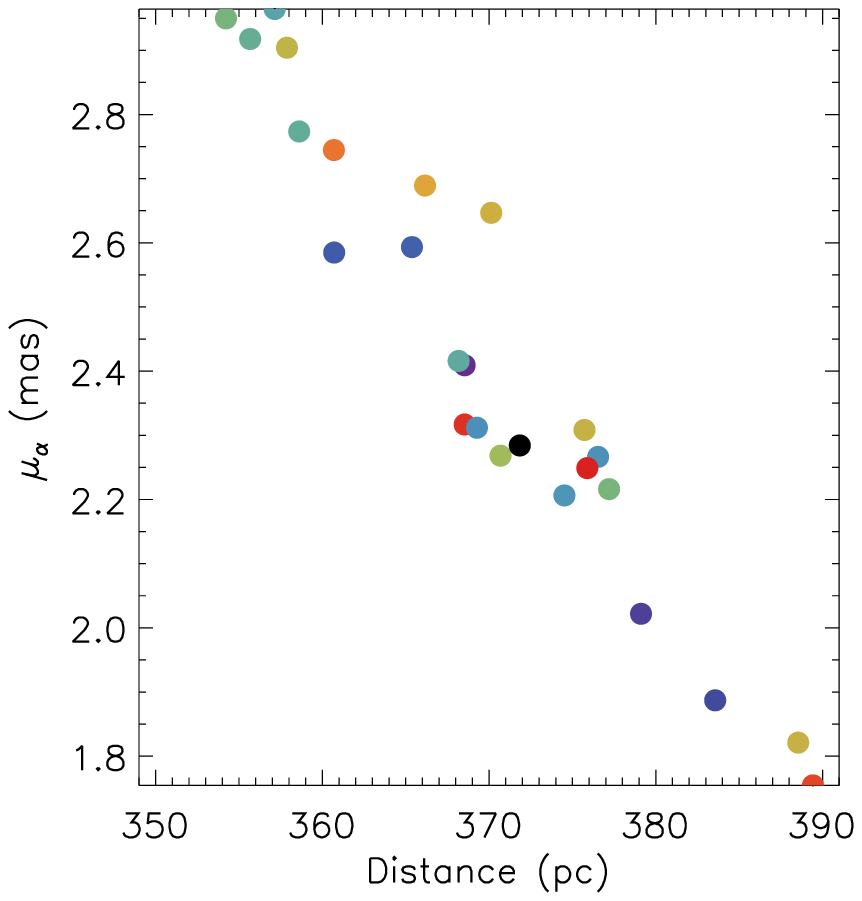}{0.15\textwidth}{}
              \fig{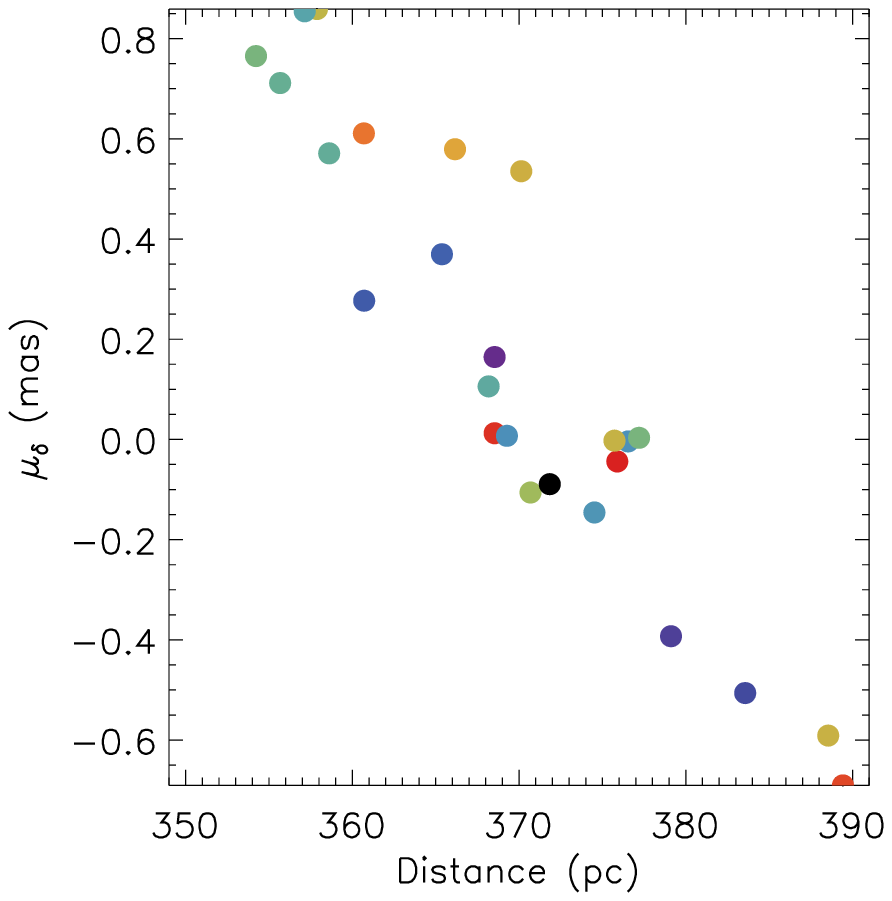}{0.16\textwidth}{}
              \fig{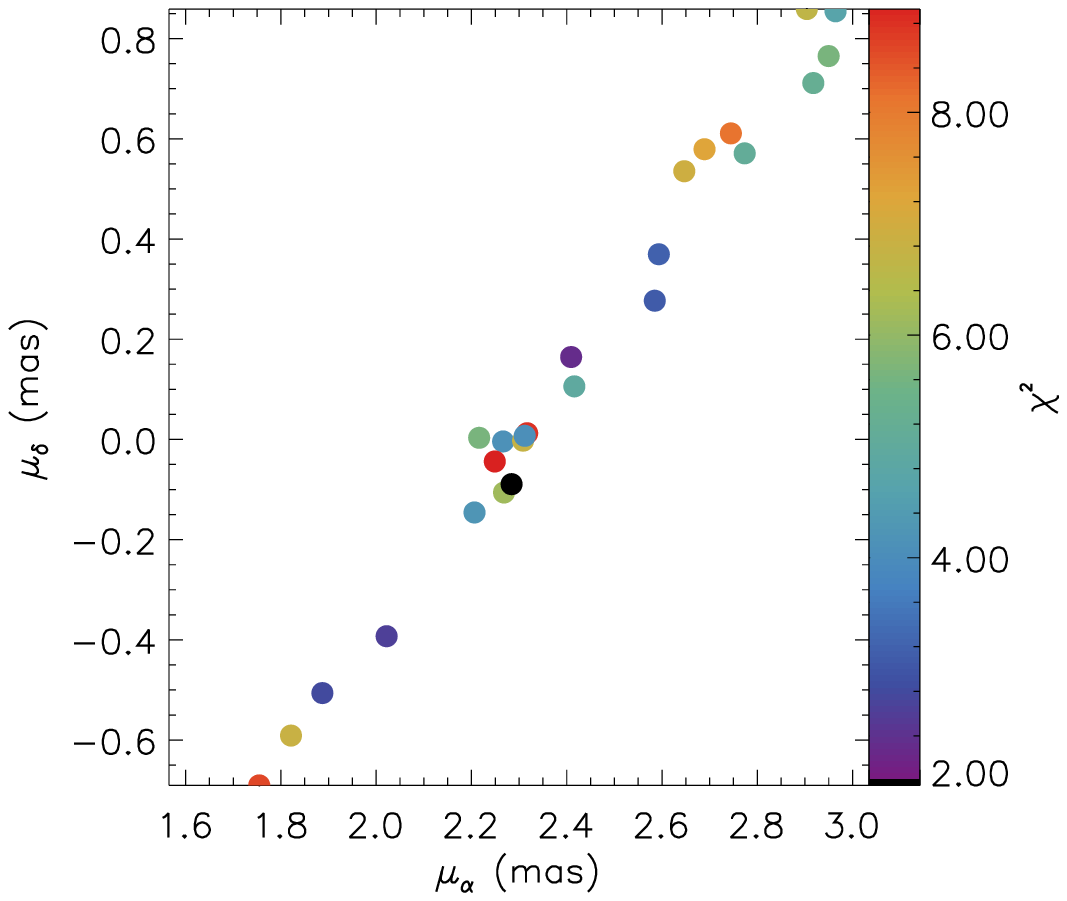}{0.19\textwidth}{}
             }
	\gridline{\fig{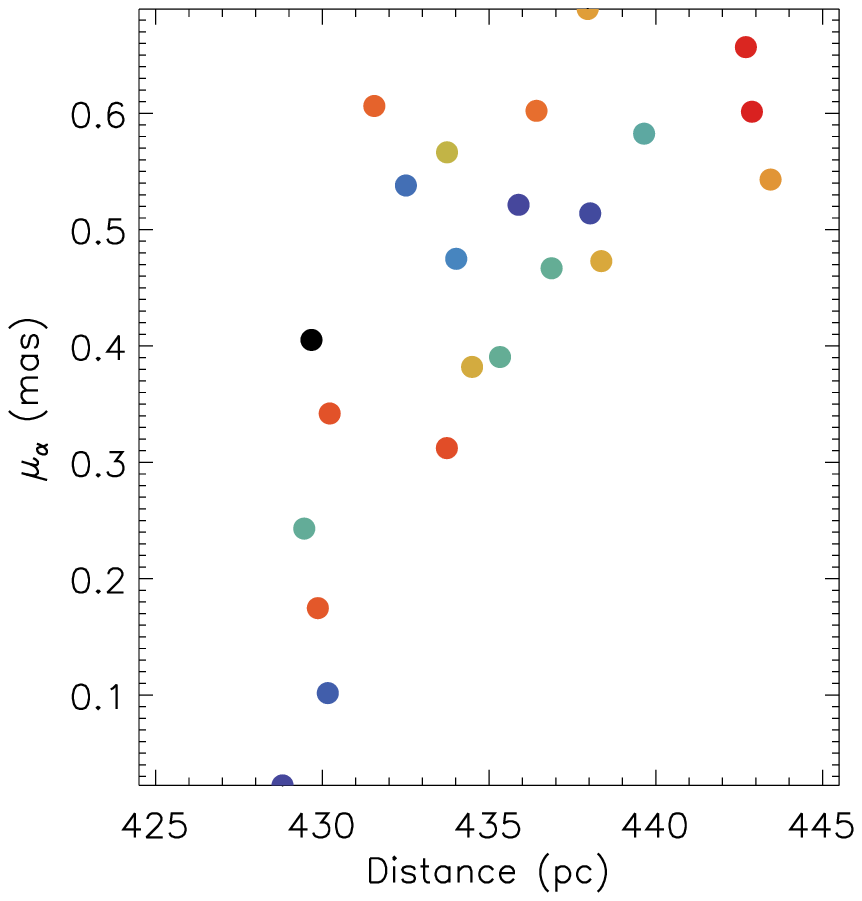}{0.15\textwidth}{}
              \fig{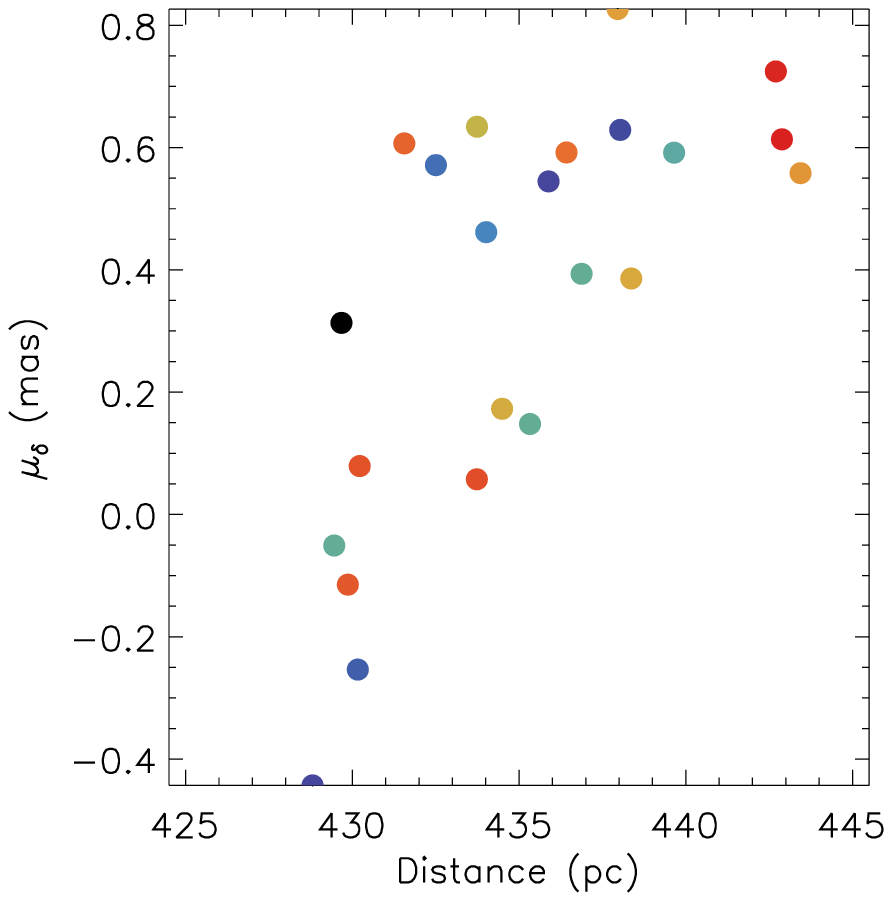}{0.16\textwidth}{}
              \fig{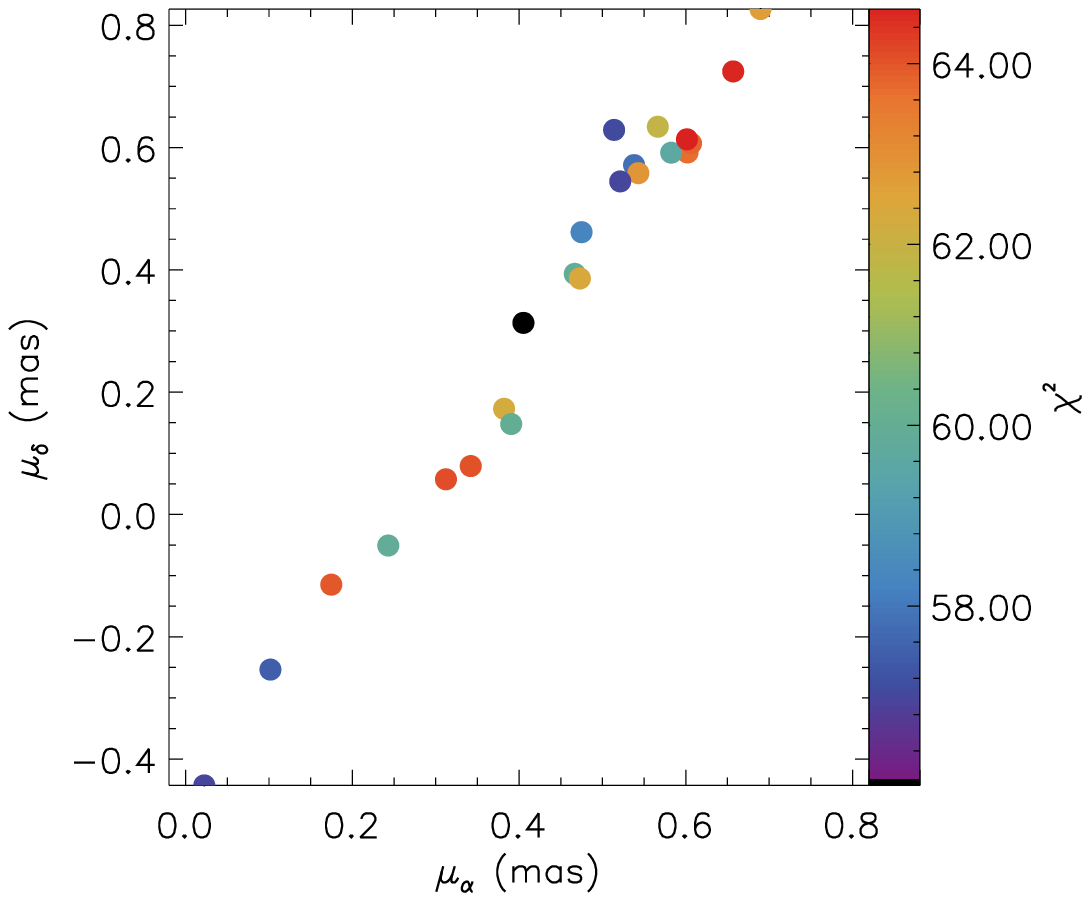}{0.19\textwidth}{}
             }
 \caption{Dependence and scatter in the fitted distance and proper motions of the astrometric binaries VLBA 4/107 (top) and VLBA 61/62 (bottom) in the different realization of the fit up to the reduced $\chi^2$ of 10 and 65 respectively. \label{fig:bin}}
\end{figure}

The fitting code is not optimized for determining several of the Keplerian parameters in a robust manner, as MPFIT is not a global optimizer, and it can get stuck in the local minima if the initial guesses for the parameters are not optimal. For this reason we explore a parameter grid of the initial guesses of $P$ in steps of 0.2 years, $e$ in steps of 0.1, $T_P$ in steps of $P$/12 and $\omega$ in steps of 30$^\circ$. The final values of these parameters can be fine-tuned by the code, and the remaining orbital and parallactic parameters are fitted directly. The uncertainties are determined from the combination of the uncertainties produced by the fit as well as the scatter in the fits from the various initial guesses for the parameters in the grid. Due to a limited number of available epochs, there is a minor dependency in some of the fitted parameters (e.g. parallax and proper motions) between the different realizations of the fits, although the exact trend between the fitted distance and proper motions may be more or less systematic depending on any number of factors, e.g. number of epochs monitored, inclination, or any other physical properties of the system (Figure \ref{fig:bin}), although the uncertainties in the parameters do take the range of scatter into account. The results of these fits are typically comparable within 1$\sigma$ to the fits produced by the \textit{Binary Star Combined Solution Package} from \citet{2001Gudehus}. The comparison of these two implementations of the binary fitting algorithm is discussed in \citet{oph2}

\item Binaries with orbital period smaller than the time between the consecutive epochs of observation (e.g. $<$6 months). As the stars in these compact binaries should not move far from its center of mass, the overall fit should approximate that of a single star, but with somewhat larger uncertainties in the parallax due to the random sampling of the positions of the star in its orbit, and the effect becomes minimal with a sufficiently large number of epochs. If the star in question belongs to a known spectroscopic binary with a constrained orbit, then by superimposing the orbit onto the parallactic motion it is possible to minimize this offset and determine a more reliable distance through fitting of the inclination $i$ and the longitude of the ascending node $\Omega$ for the system. In this case $e$, $P$, $a_1\sin i$, $T_P$, $\omega$, and $q$ are held fixed to the known values from the spectroscopic orbital solutions.

\end{enumerate}

\section{Discussion}

\begin{figure*}
	\gridline{\fig{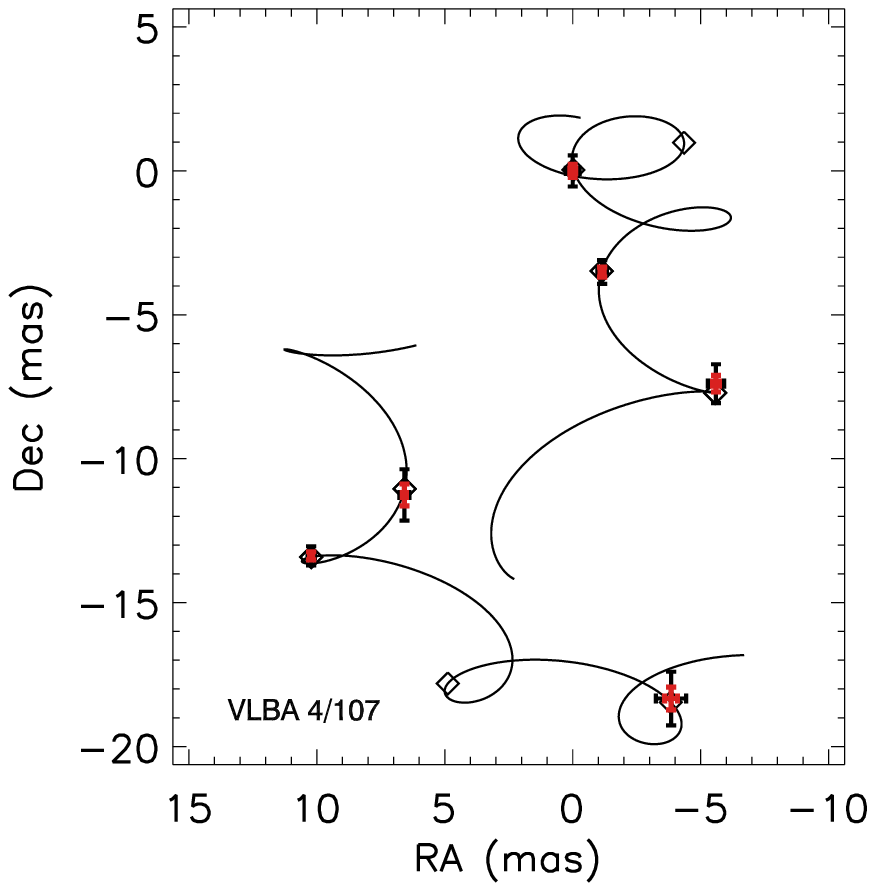}{0.17\textwidth}{}
              \fig{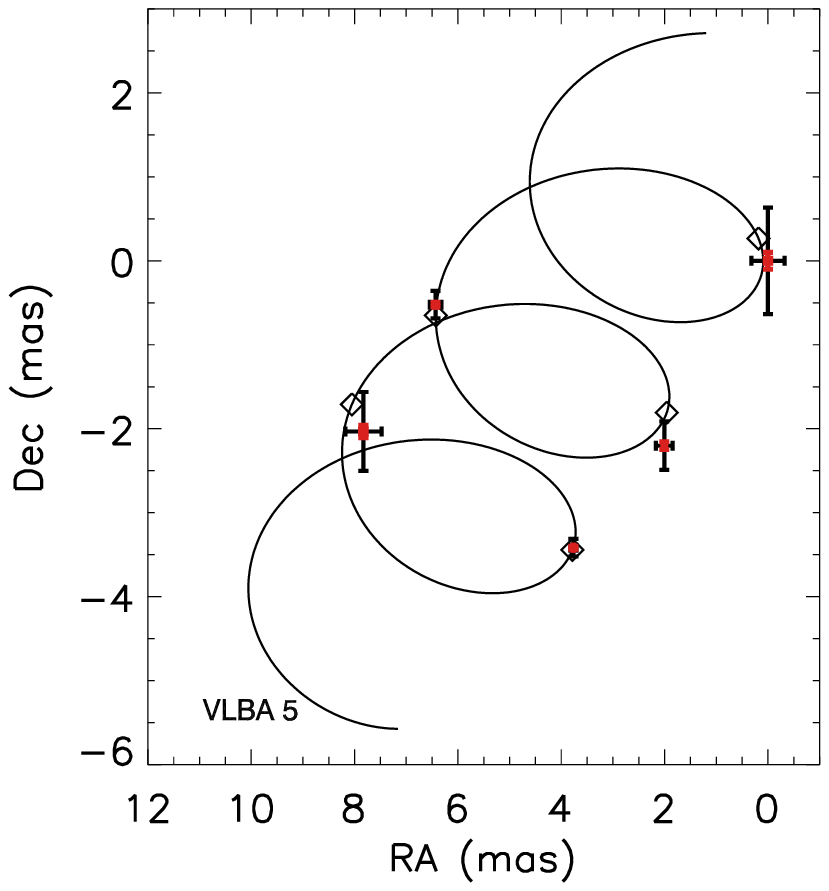}{0.16\textwidth}{}
              \fig{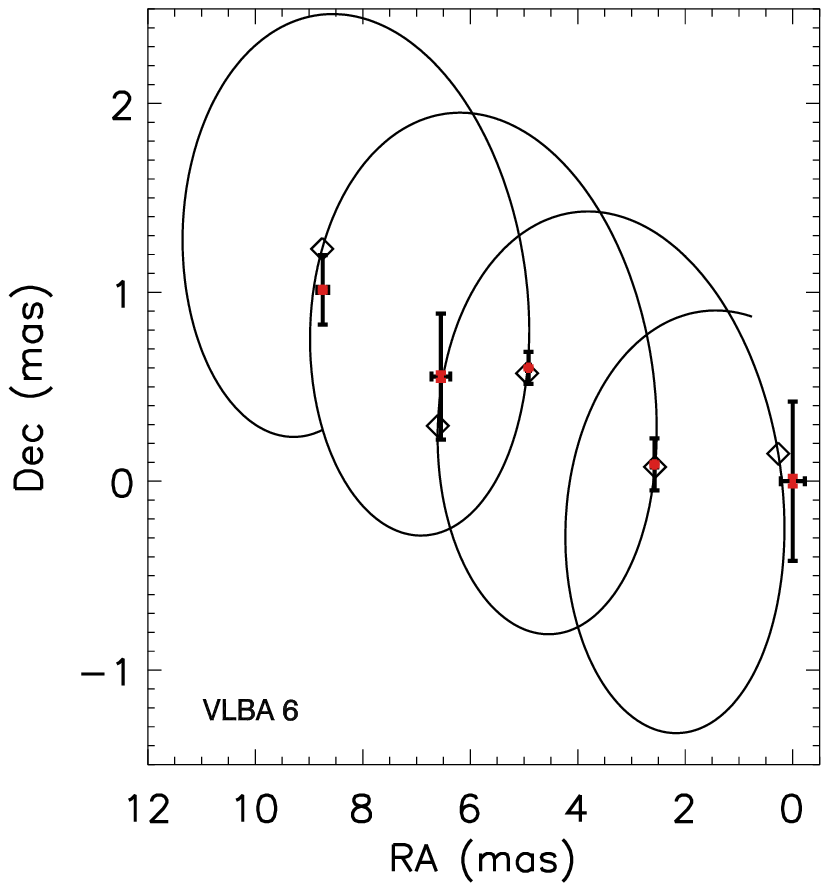}{0.16\textwidth}{}
              \fig{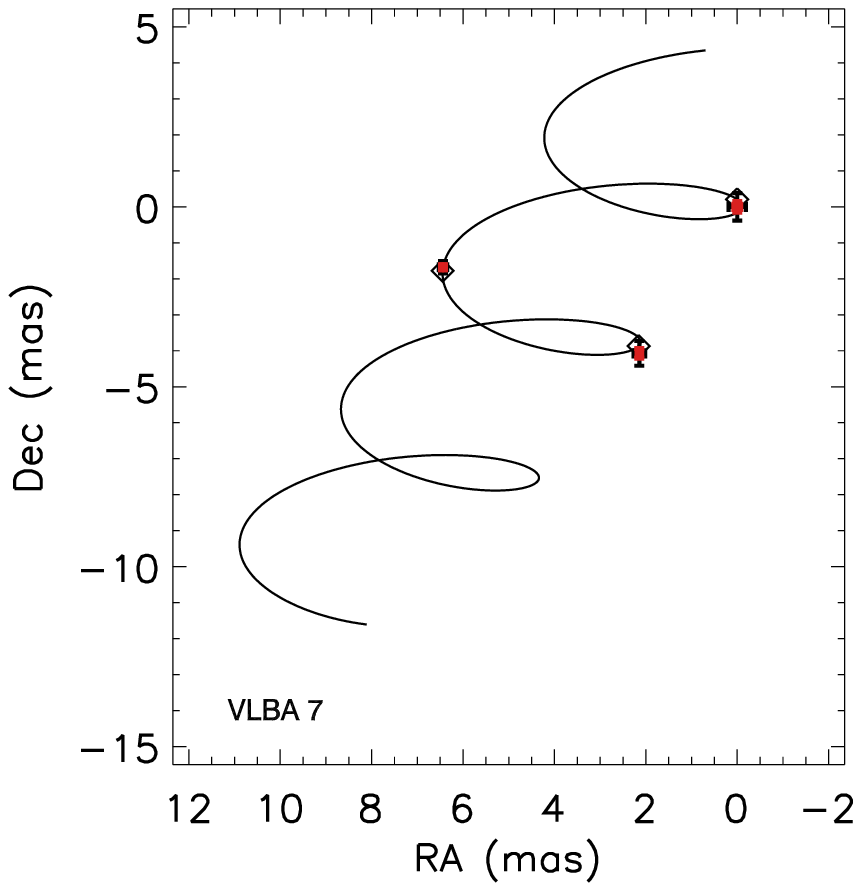}{0.165\textwidth}{}
			  \fig{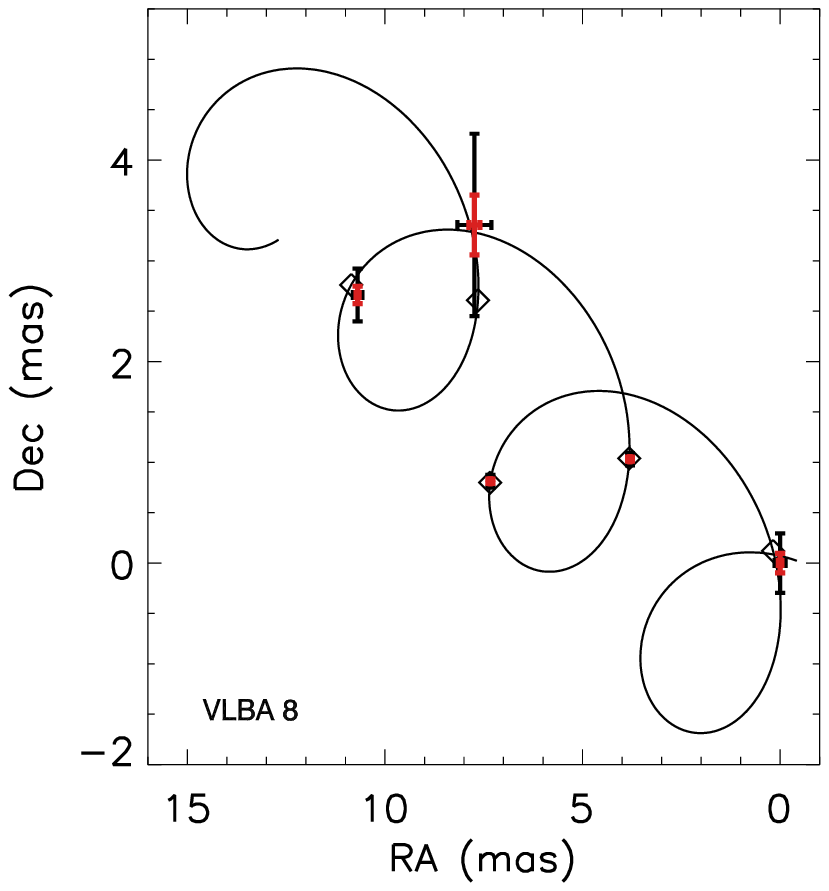}{0.16\textwidth}{}
              \fig{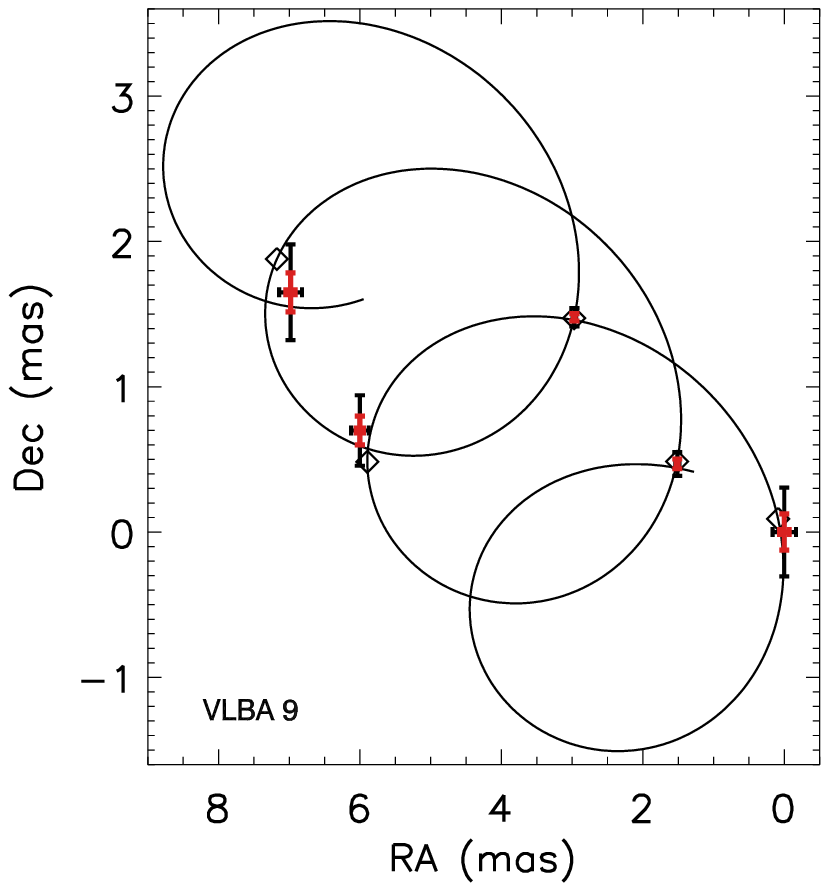}{0.16\textwidth}{}
             }
	\gridline{
              \fig{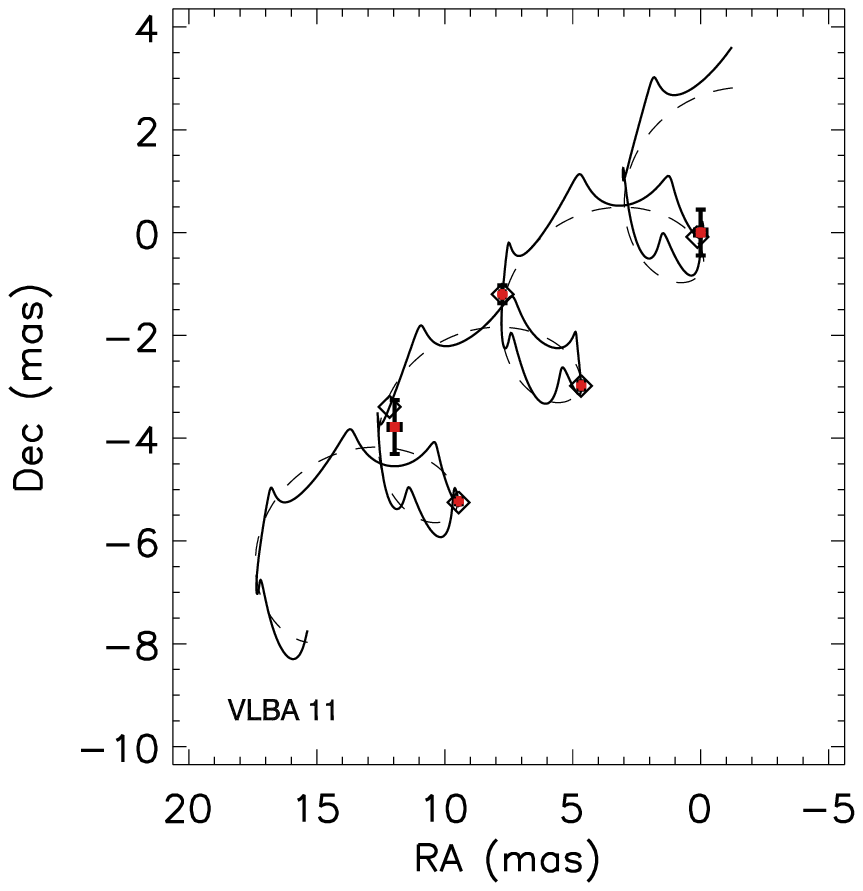}{0.17\textwidth}{}
              \fig{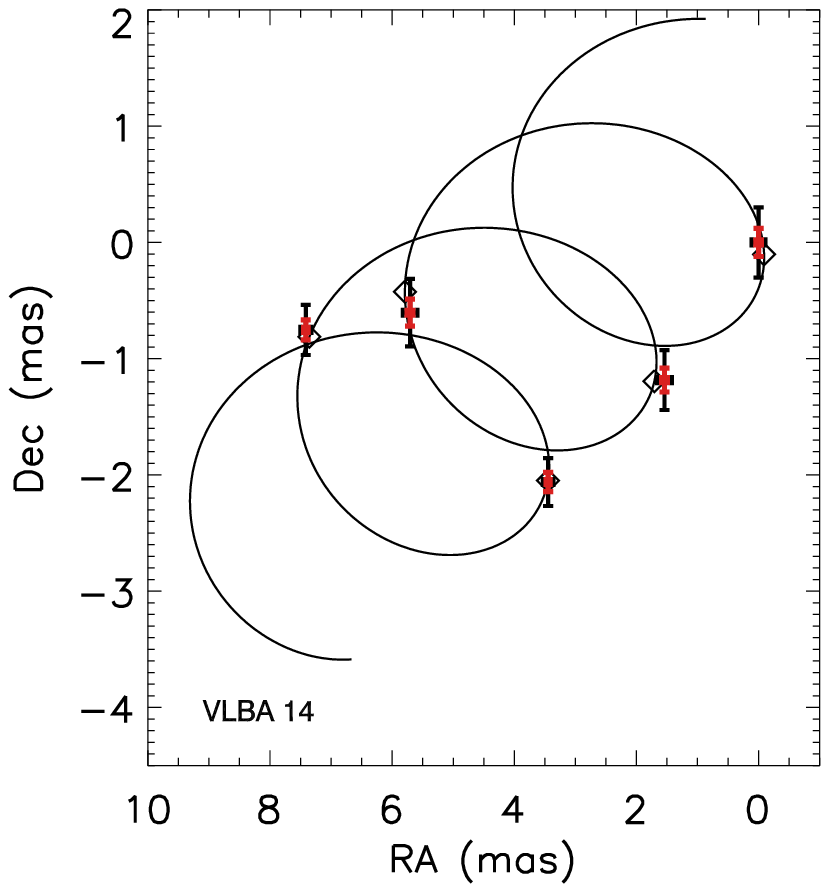}{0.16\textwidth}{}
              \fig{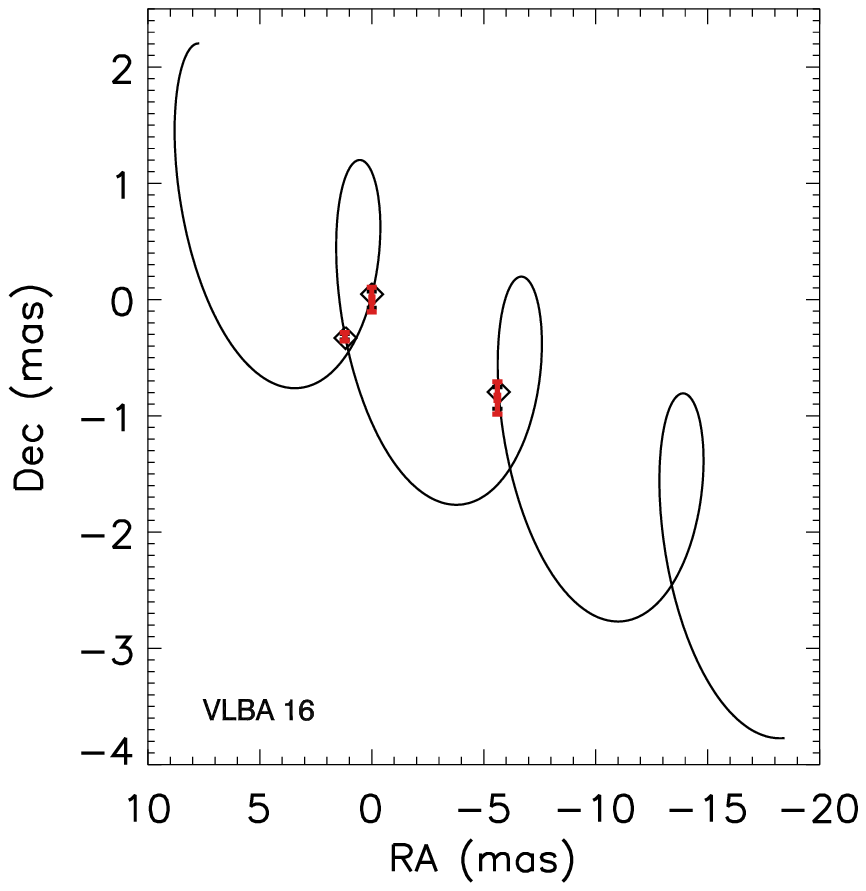}{0.17\textwidth}{}
              \fig{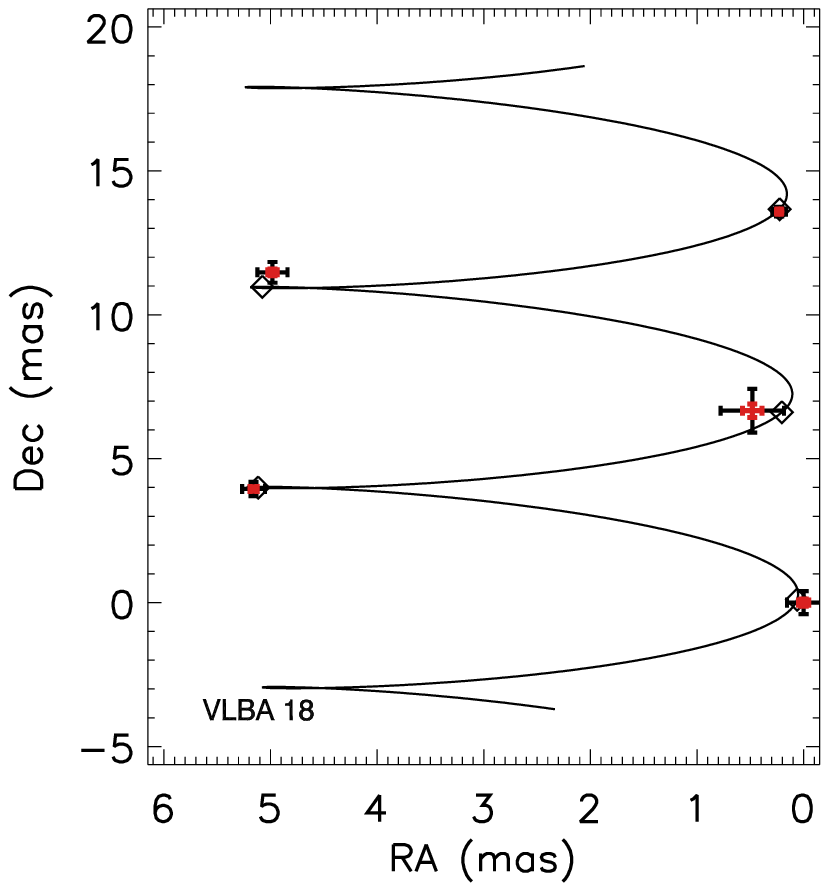}{0.16\textwidth}{}
              \fig{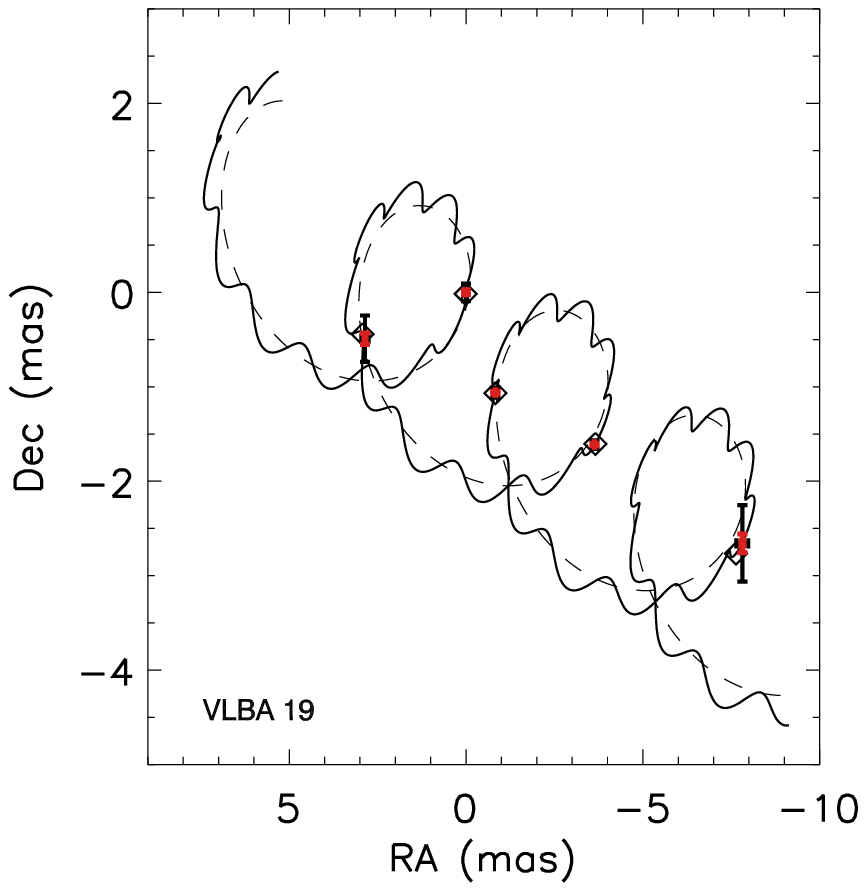}{0.17\textwidth}{}
              \fig{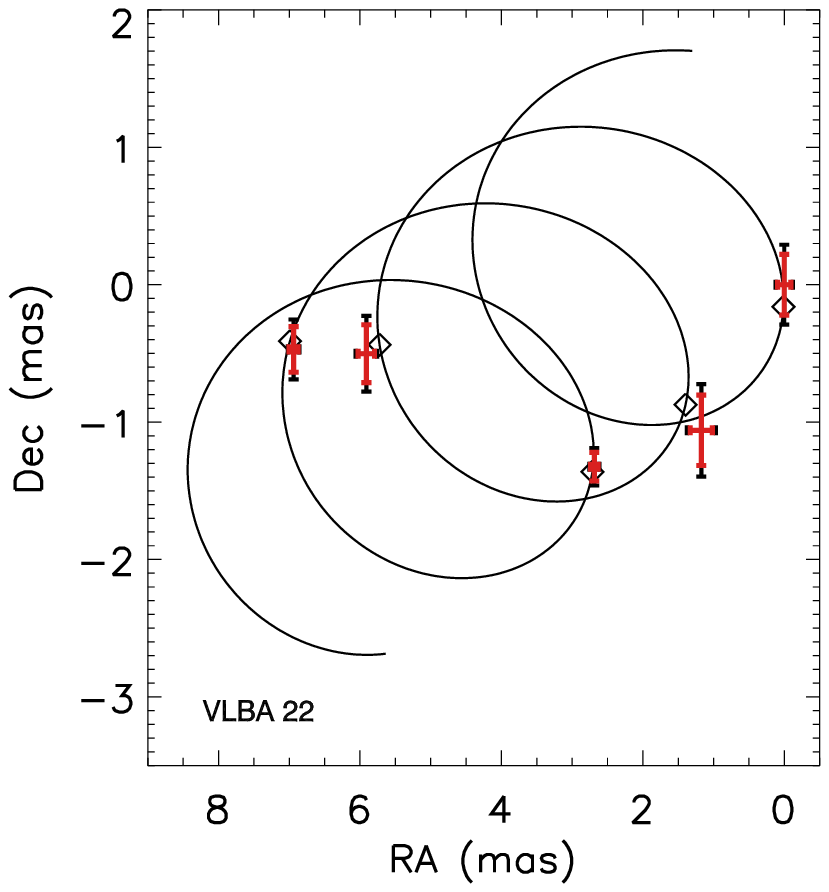}{0.16\textwidth}{}
             }
	\gridline{
              \fig{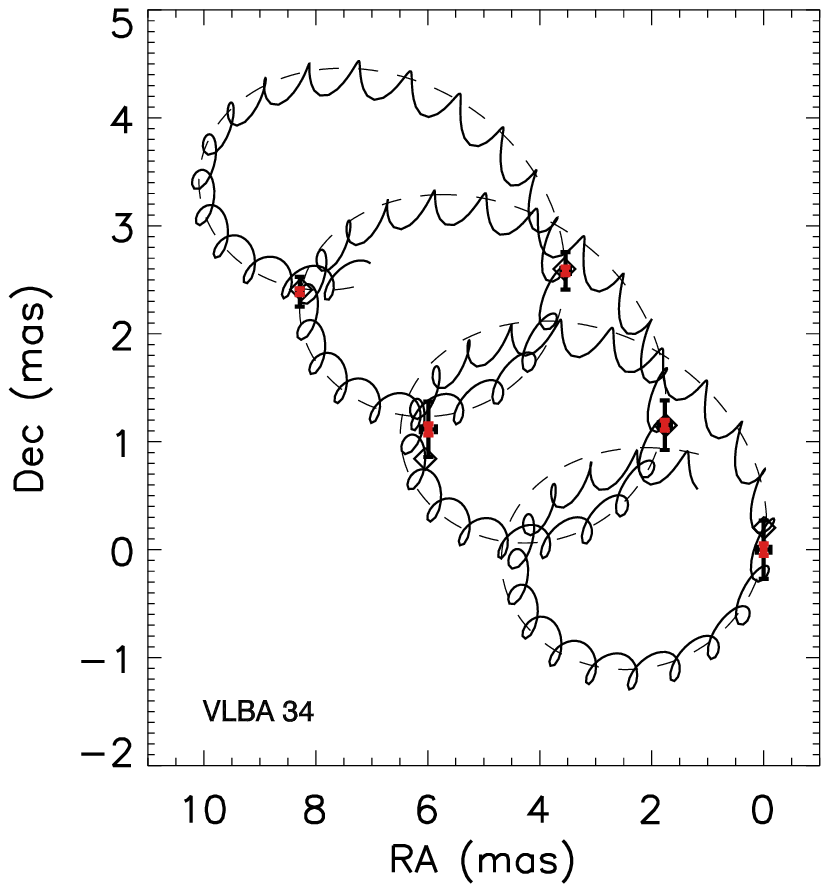}{0.16\textwidth}{}
              \fig{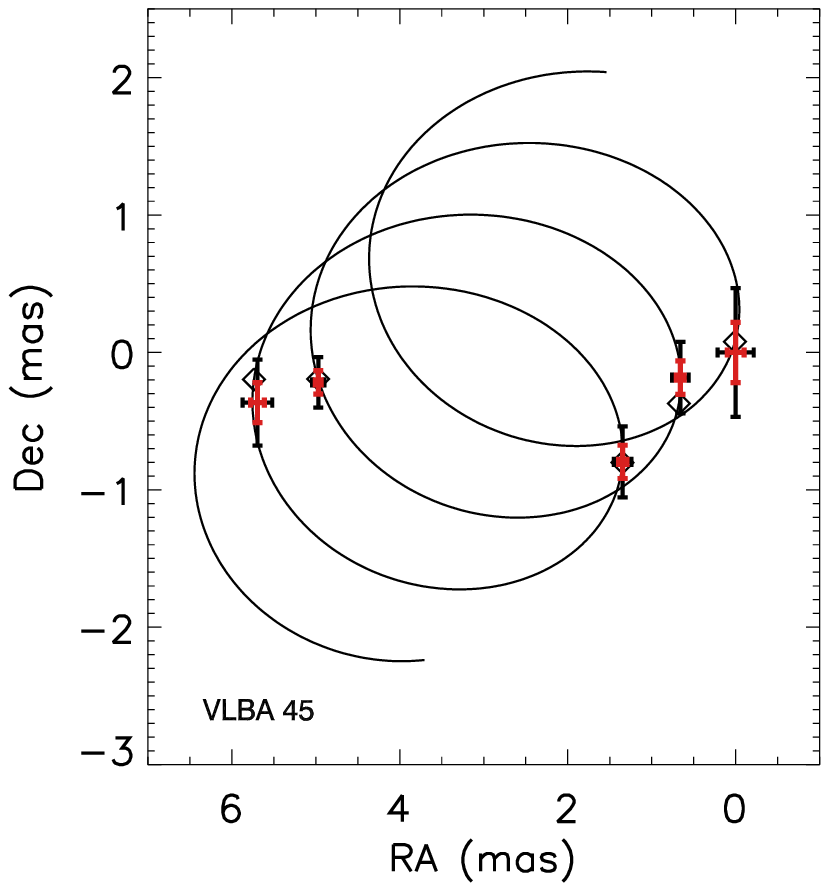}{0.16\textwidth}{}
              \fig{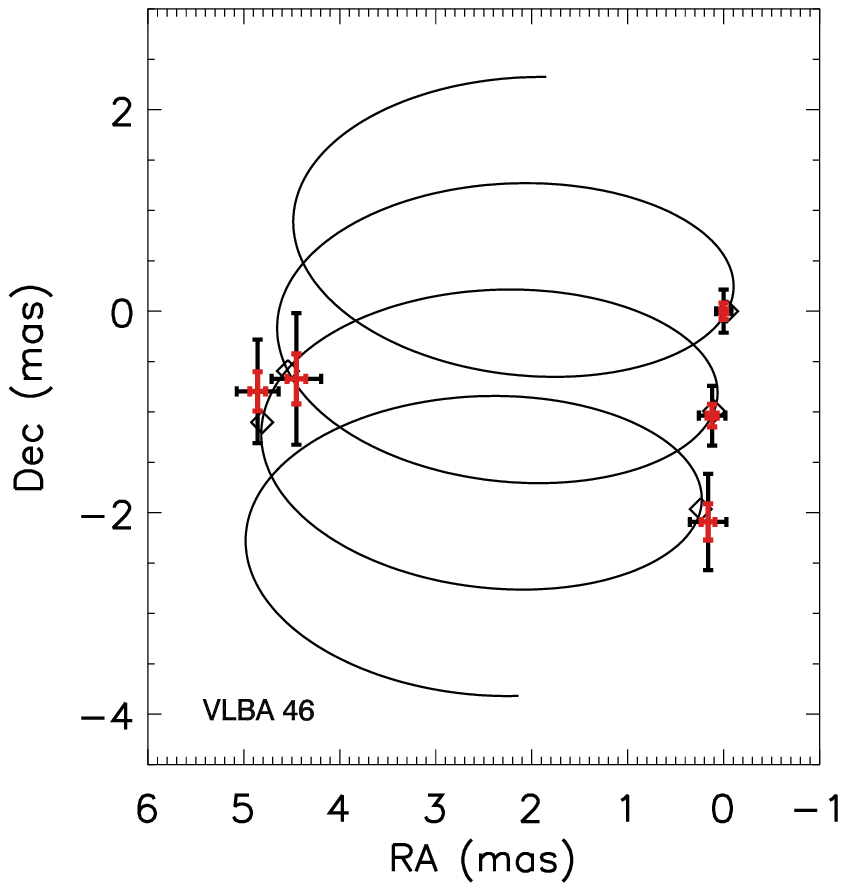}{0.165\textwidth}{}
              \fig{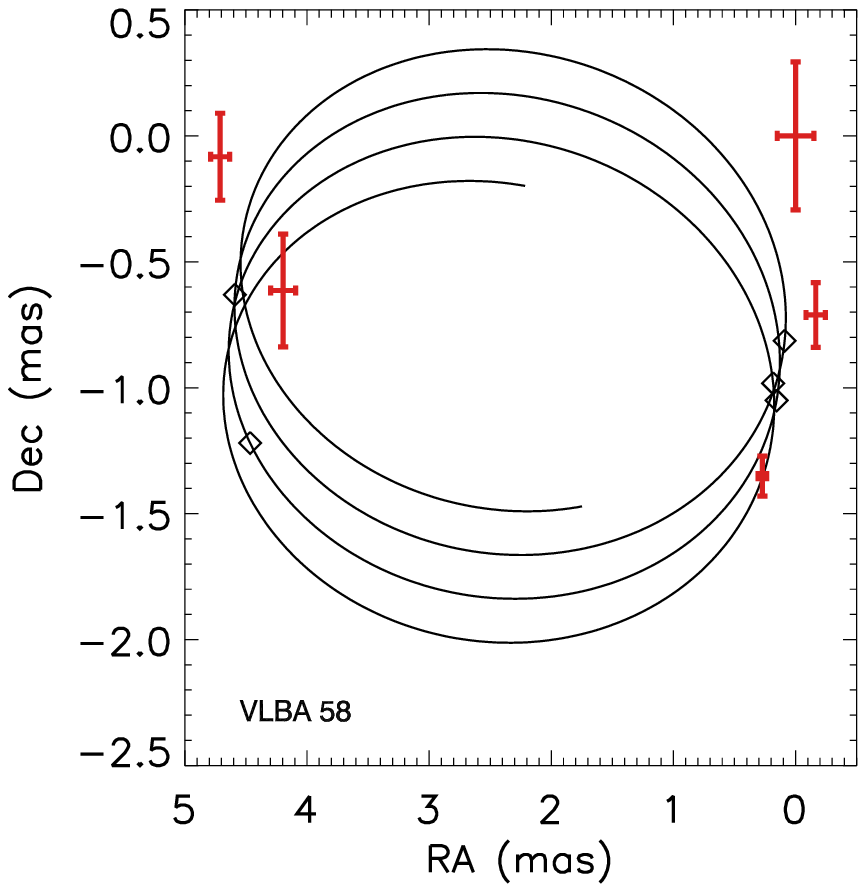}{0.17\textwidth}{}
              \fig{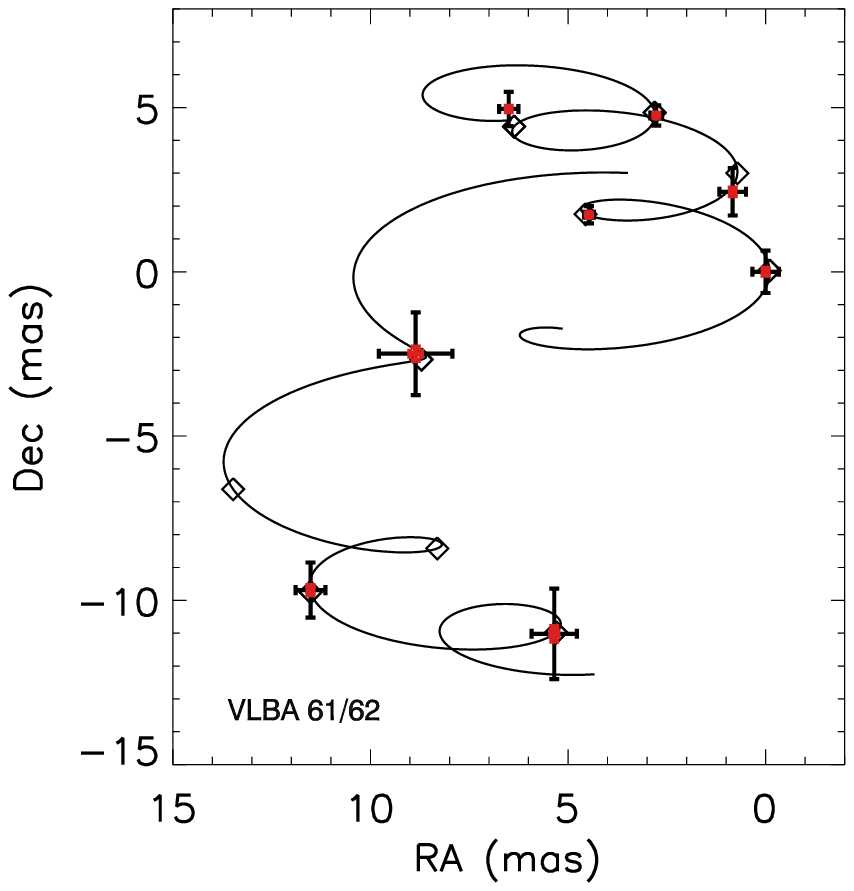}{0.165\textwidth}{}
              \fig{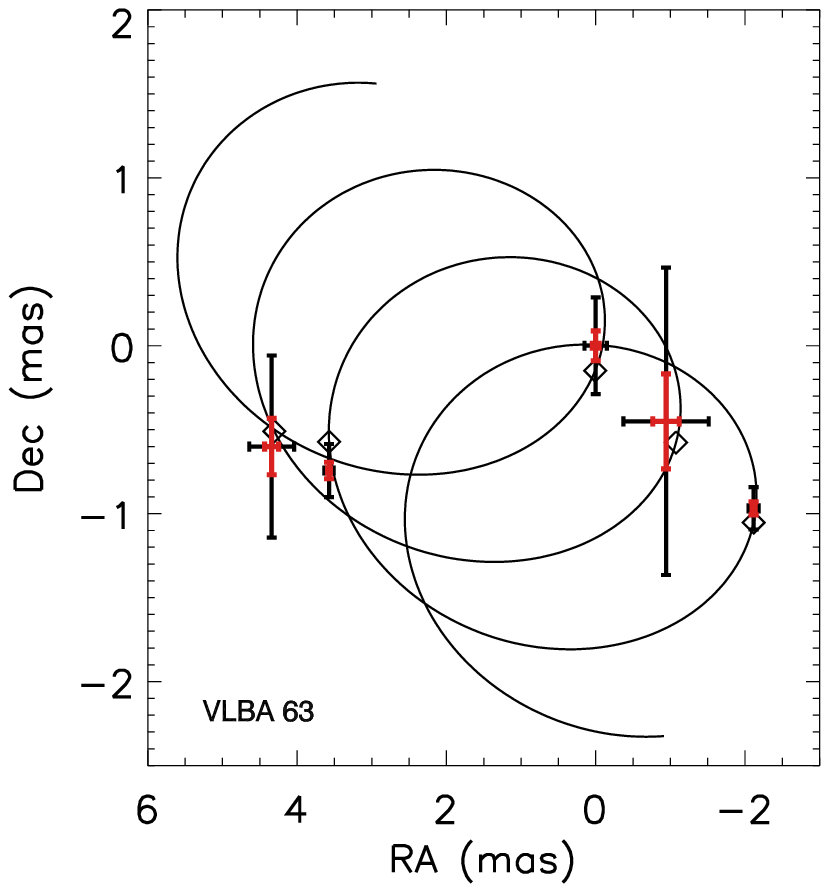}{0.16\textwidth}{}
             }
	\gridline{
              \fig{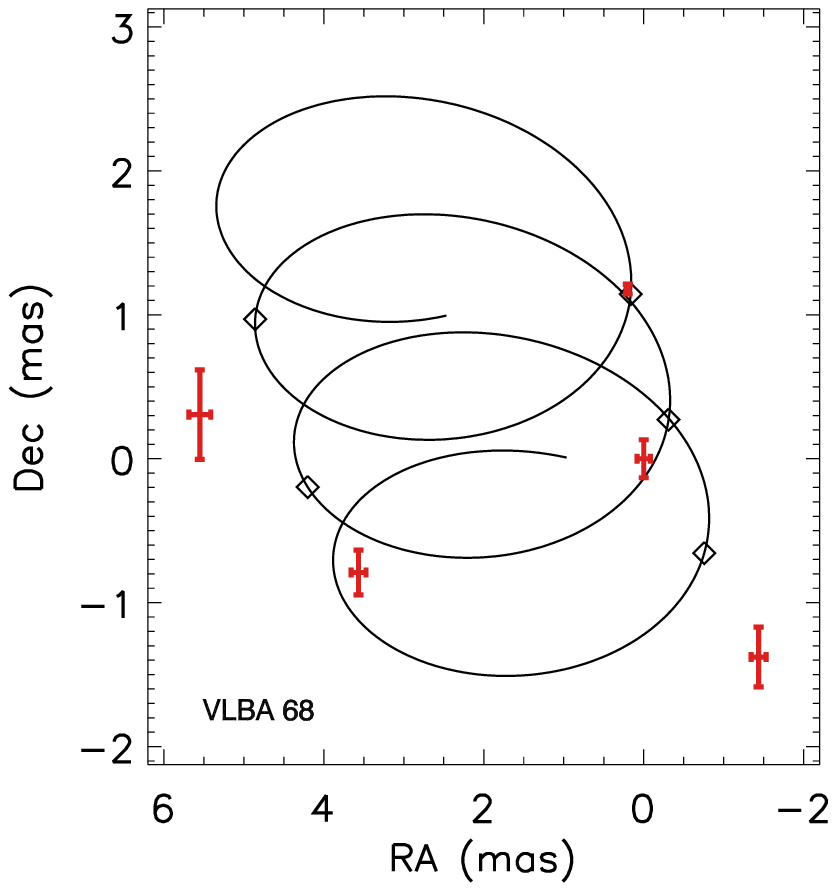}{0.165\textwidth}{}
              \fig{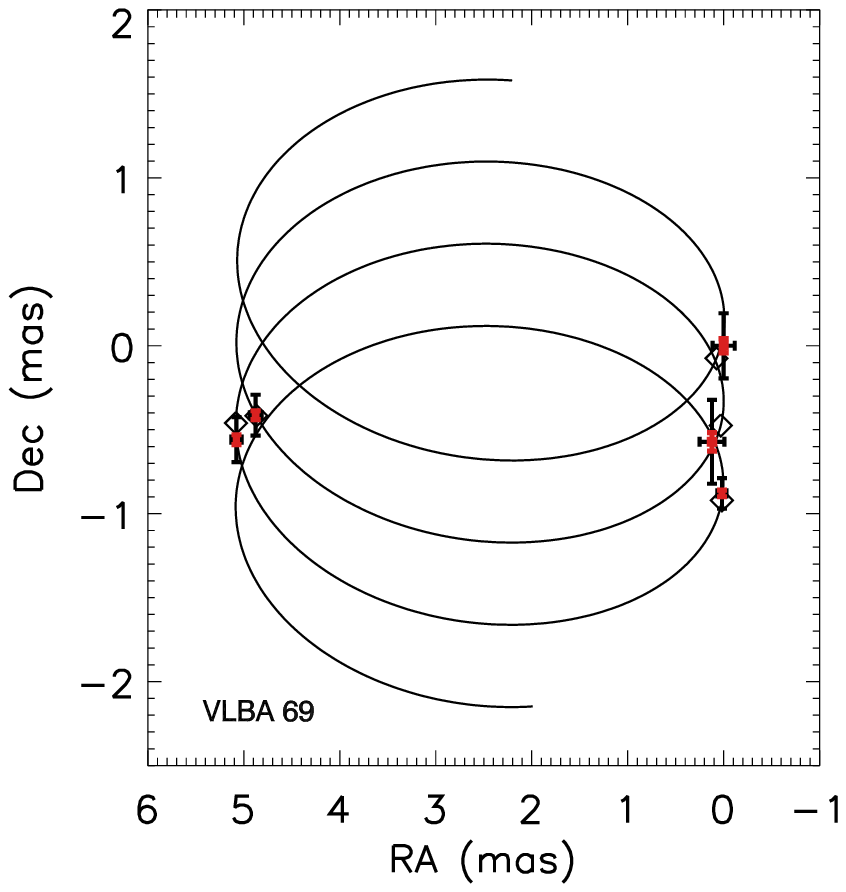}{0.165\textwidth}{}
              \fig{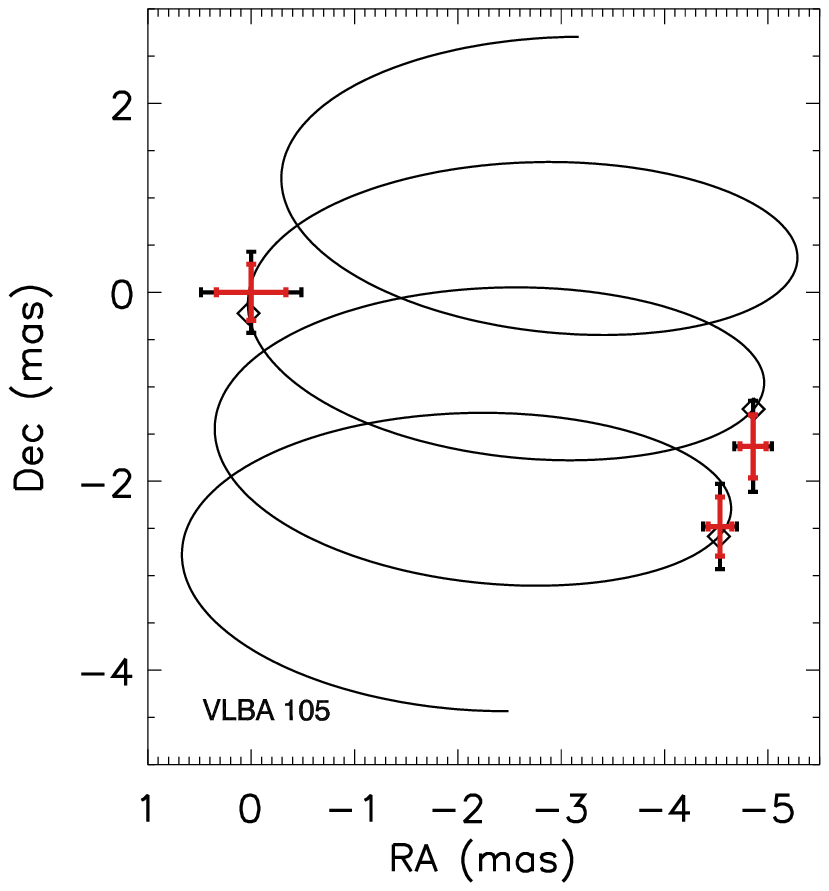}{0.16\textwidth}{}
              \fig{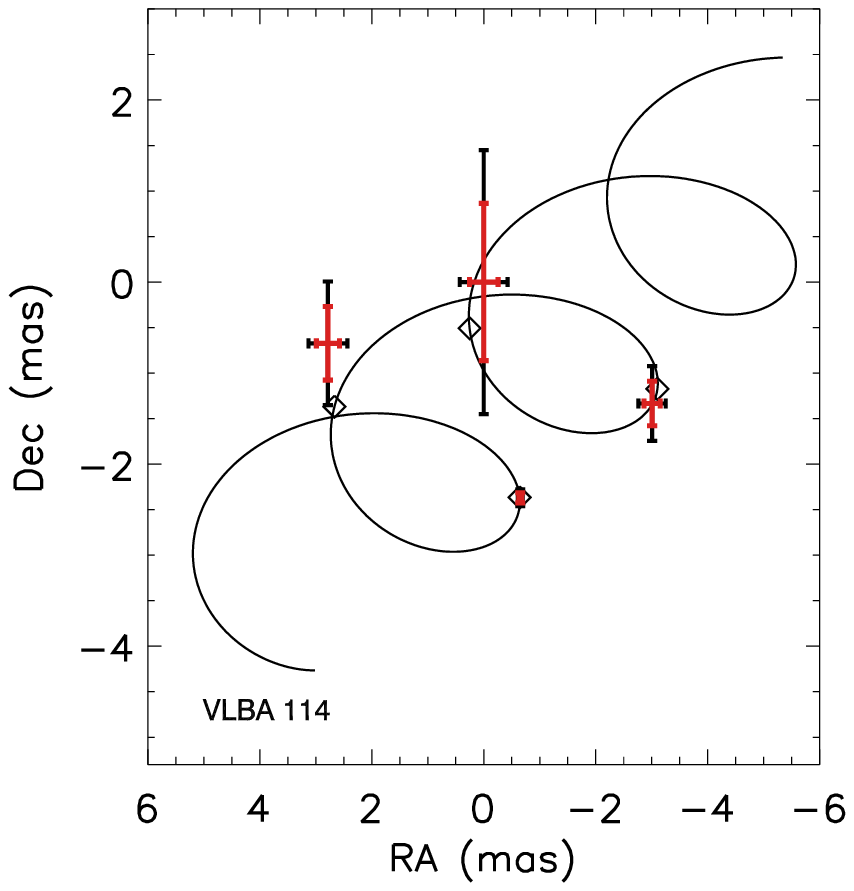}{0.165\textwidth}{}
              \fig{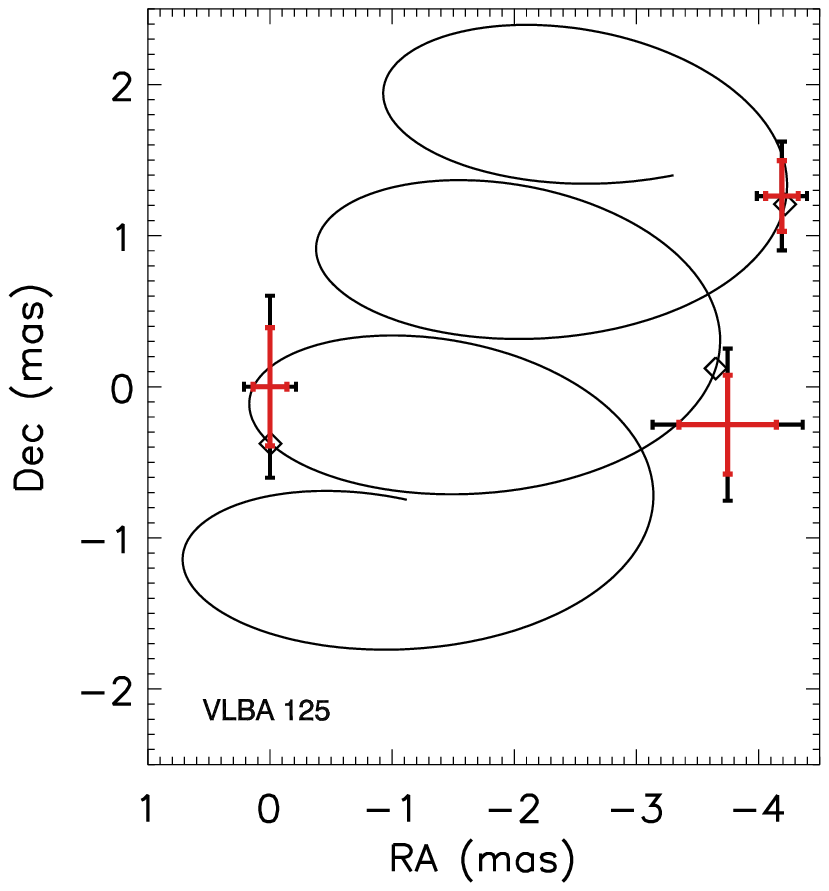}{0.16\textwidth}{}
              \fig{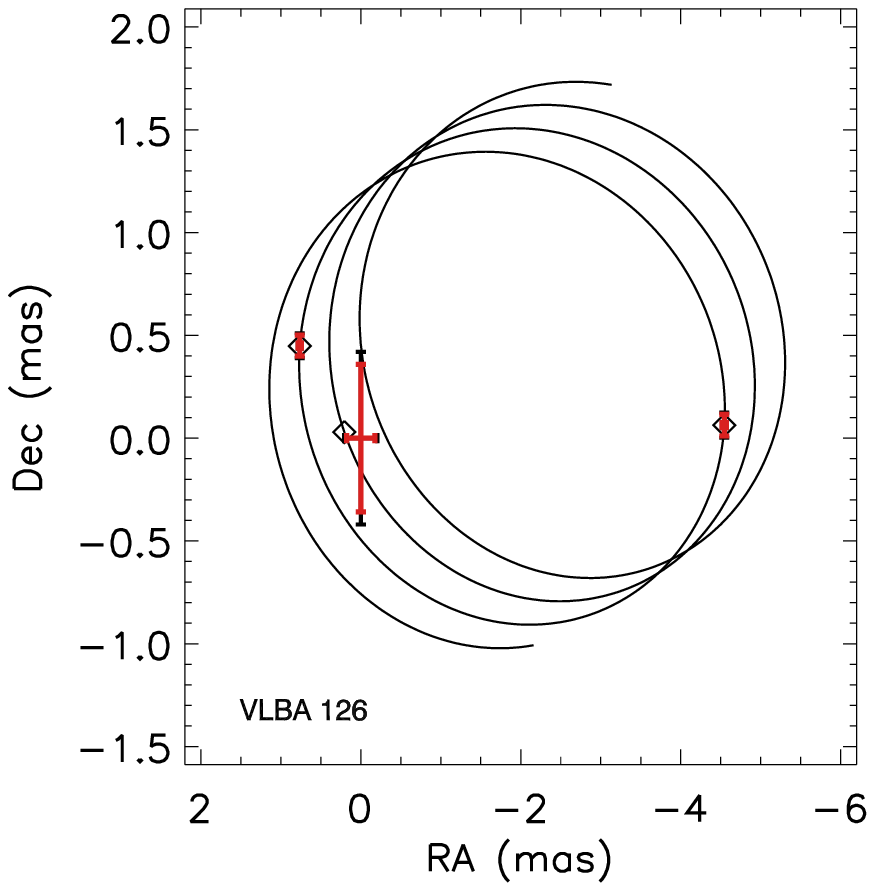}{0.17\textwidth}{}
             }
	\gridline{
              \fig{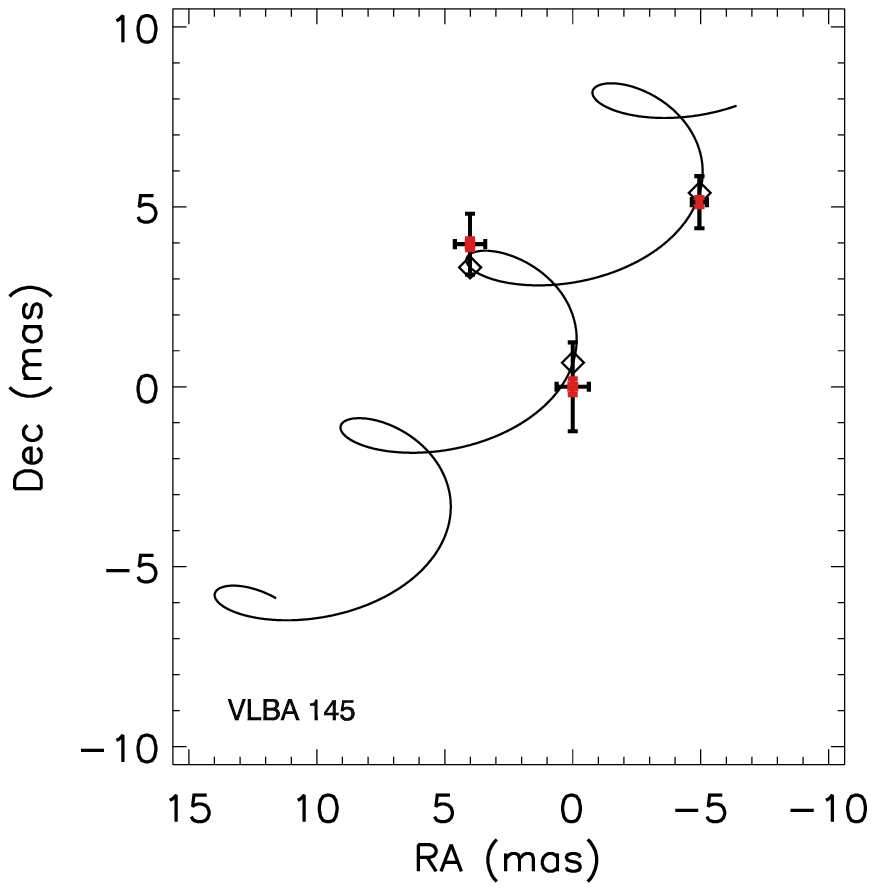}{0.17\textwidth}{}
              \fig{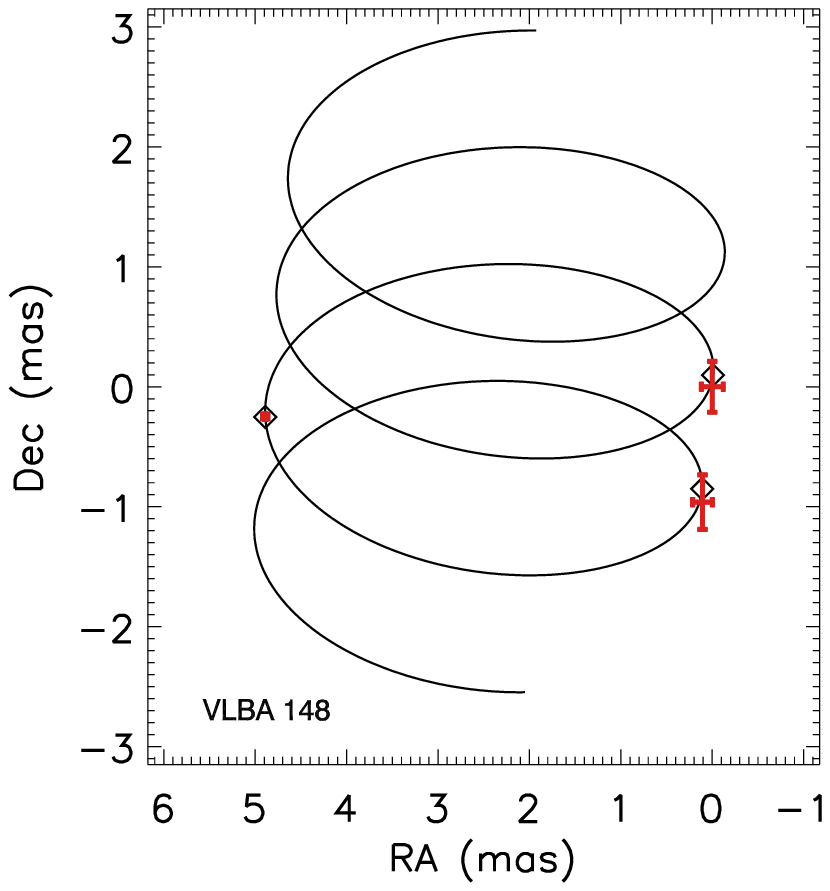}{0.16\textwidth}{}
             }
 \caption{Best fits for the data. Red error bars show astrometric uncertainties produced by JMFIT, black error bars show uncertainties scaled by the value needed to achieve $\chi^2=1$. Diamonds show expected position of a star at the time of the observations based on this fit. Dashed lines indicates the fit assuming a single star for spectroscopic binaries.\label{fig:par}}
\end{figure*}

Twenty stars have been detected towards the ONC, and we present distance solutions to fifteen of them . Six of these stars belong to the Trapezium cluster. Two stars have been detected towards L1641, three towards NGC 2068, one towards $\sigma$ Ori. Seven stars have been detected towards NGC 2024, and we present preliminary distance solutions to five of them. These solutions are presented in Table \ref{tab:par} and discussed in Appendices B for ONC and C for the remaining regions. The individual fits are presented in Figure \ref{fig:par}.

Three stars are found to be astrometric binaries with both components detected - VLBA 4/107, 27/28, and 61/62. We presently can fit for the orbital motion of VLBA 4/107 and 61/62, the parameters of which are presented in Table \ref{tab:bin}. Another two, VLBA 58 and 68, are also identified as belonging to a multiple systems with intermediate period on the basis of their astrometry, although only a single star has been detected. Additionally, VLBA 125, 126, and 145 have been detected in only three epochs, but they also show possible signatures of multiplicity. Six stars, VLBA 6, 9, 11, 19, 27, and 34, are known spectroscopic binaries with short periods. The distance solution of VLBA 11, 19, and 34 incorporates the orbital motion. The parameters of the orbit are presented in Table \ref{tab:spbin}. Another four stars, VLBA 4/107, 8, 11, and 27/28, also have either known or identified long period companions in addition to the aforementioned nearby companions (with exception of VLBA 8, they belong to a high order multiple systems).

\subsection{Revised distance to the Trapezium}

Out of six stars observed in the Trapezium, four have been previously observed by MR, although they have incorporated only three of them into the distance solution for the cluster. A simultaneous parallactic fit of all the stars found towards the Trapezium in this program results in a distance of 383$\pm$3 pc. This fit is produced by fitting the equations of motions of all the stars at the same time with a single distance but different proper motions for all stars. The result of this fit is also identical to the weighted average of the individual measurements. The weighted average distance of all the stars in the ONC is 388$\pm$2 pc. These values do not include possible systematic effects due to the phase gradient (see bellow). This is discrepant with the distance of 414$\pm$7 pc obtained by MR by 3$\sigma$, or $\sim 0.2$ mas in parallax. It is possible that there is a systematic offset of such magnitude between different epochs that cannot be reproduced during fitting by merely scaling the positional uncertainties until the $\chi^2$ of the fit is equal to 1, therefore it is possible that the formal uncertainties are somewhat underestimated. And, since all of these stars are observed in a single field, any systematic offset that is applied to the coordinates of the center of the field will be propagated to the positions of the sources, and the parallaxes and proper motions could be affected accordingly. However performing a fit with a reduced number of epochs in either work done by MR or this work offers no reconciliation, therefore the effect is not dominated by a pointing error in any single epoch.

Low frequency radio observations could be affected by the dispersive delay \citep[e.g.][]{2014Reid} that is difficult to calibrate resulting in a phase gradient across the sky, producing a slight offset in the absolute positions of the targeted objects. This effect becomes large the further the object is from the primary calibrator. It does "average out" with the sufficiently large number of epochs; nonetheless, some of it can propagate to the parallax estimation. This effect can lead to somewhat different distances when using different calibrators as a reference for the absolute coordinates of the targeted sources.

To estimate the strength of this phase gradient, we compare the distances of the Trapezium sources with the coordinates referenced to the observed positions of the secondary calibrators (Figure \ref{fig:all}). The primary calibrator for the field is J0539-0514, which is located 1.2$^\circ$ away from the targets. The simultaneous fit gives 394$\pm$3 pc when all the coordinates are referenced to J0529-0519 (1.3$^\circ$ away), 375$\pm$3 pc relative to J0541-0541 (1.6$^\circ$ away), and 382$\pm$3 pc relative to J0532-0307 (2.4$^\circ$ away). As the Trapezium sources are located approximately halfway between J0539-0514 and J0529-0519, we can estimate the systematic effect of the dispersive delay on the parallax towards it to be on the order of 0.033 mas (5 pc at the distance of the ONC). This effect is consistent throughout the ONC as long as these calibrators are separated by less than 1.5$^\circ$ from the target. Referencing the coordinates to J0529-0519 tends to produce somewhat larger distances than referencing them to J0539-0514. This is also true in the cases where J0529-0519 is used as the primary calibrator (i.e. in the cases of VLBA 14, 16, 18, and 19). We add 5 pc (or 0.033 mas) in quadrature to the uncertainties in the weighted average distance to account for this systematic effect. We adopt a distance of 388$\pm$5 pc towards the ONC, including the Trapezium.

This analysis cannot be performed on the solutions obtained by MR, as they have observed only a single calibrator. On one hand, as their observations were obtained at higher frequency (8.4 GHz), they should be less susceptible to the dispersive delay. On the other hand, they have obtained fewer epochs, thus this effect is somewhat more likely to propagate to the parallax, and their primary calibrator is somewhat further away.

Some minor differences could also be attributed to the difference in the fitting routines. Both the codes used in this work and work done by MR produce fits within 1$\sigma$ of each other when applied to a particular set of positions. However, MR assume a circular orbit for the Earth and fit parallax only from $\alpha$, using $\delta$ only to fit $\mu_\delta$, whereas the code used in this work considers the effect of the parallax on both $\alpha$ and $\delta$ using the true orbit of the Earth. Fitting positions quoted in MR produces a combined distance of 406$\pm$4 pc including GMR G, or 409$\pm$3 pc excluding it.

While it is possible to make a single fit for each star utilizing the data obtained by both S07, MR and this work, the difficulty lies in the fact that the each survey utilized a different observing and calibration strategy. S07 used J0541-0541 as a primary calibrator and J0529-0519 as the secondary calibrator. MR used J0541-0541 as the primary calibrator as well as the geodetic sources. This work used J0539-0514 as the primary calibrator with three secondary calibrators and geodetic sources. The absolute positions of the calibrators that are found in common between these works are assumed to be somewhat different, on up to a $\sim 1$ mas level. All of these factors introduce an offset in the absolute positions between these works that is not found in observing the sources repeatedly with the same calibration strategy. While it is possible to calibrate the magnitude of this offset, the exact fit is strongly dependent on the manner of calibration, therefore it is best to treat the data obtained by different projects independently. However, as both this work and S07 have a larger number of observations that the work done by MR, the fit utilizing all the positions does seem to favor a significantly closer distance than the one found by MR.

Multiplicity is another possible culprit of the difference between the fits. At least two stars observed by MR are used in their distance estimate are spectroscopic binaries, which makes distance solutions produced by them to be inherently more uncertain. We analyzed the effect of these orbits could have on the distance. The effect on distance of GMR 12 is within 1 pc. No orbital solution currently exists for GMR F. For of the remaining sources observed by this program and identified as spectroscopic binaries, the effect varies between 3 to 18 pc. While it is possible that it can contribute to the difference between these two works, it is unlikely that it could explain the systematic nature of the offset. Although it must be noted, that the orbital motion parameters can be greatly affected by any systematic offsets in the data, particularly when the orbits cannot be fitted a priori.

Nonetheless, there is sufficient evidence to suggest that the ONC is located closer than has been previously estimated by MR. Systematic offsets would affect each pointing differently. Therefore, a larger number of fields with larger number of epochs and a larger number of sources overall through the ONC implies that the overall effect of the systematic offsets, if it is present, would be more noticeable in this work than in the work done by MR. The consistency in both the fitted distance and the proper motions estimates of GMR A between this work and that done by S07 suggests that our results are reproducible, which would also be less likely if there was a significant positional offset in our data. Finally, the consistency in distance towards GMR G between this work and that done by MR is surprising, given the lack of consistency between other sources.

Finally, there exists the curious case of the NGC 1980 cluster which is located to the south of Trapezium in the vicinity of $\iota$ Ori. This cluster has a somewhat older population of stars compared to the rest of ONC \citep[4-5 Myr][]{2012alves}. Unfortunately we do not detect any YSOs towards it, but it is though to be located at the distance of 380 pc obtained through pre-main sequence fitting \citep{2014bouy}. This is why this cluster was thought to exist in the foreground of ONC, as a separate entity. Nonetheless, the kinematics of NGC 1980 do not show any unique features not present in ONC, and, in fact, the velocity dispersion towards it is the smallest of any other region found towards ONC \citep{2016dario,2016kounkela}. If we assume a significantly closer distance towards the Trapezium and the ONC than what was previously assumed by MR then NGC 1980 should not be considered a foreground cluster, but rather an integral part of the ONC.

\subsection{Structure of the Orion Complex}

\begin{figure}
\epsscale{1}
\plotone{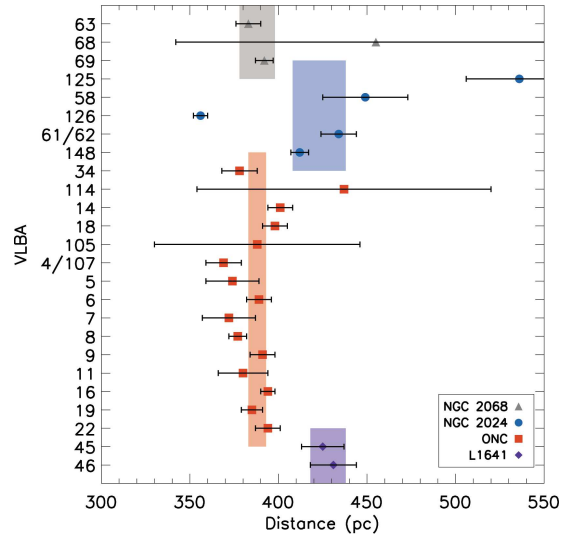} 
\caption{Measured distances to the individual stars in the four clusters, sorted according to their $\delta$. The averages for each cluster are shown with semi-transparent rectangles. \label{fig:bar}}
\end{figure}

\begin{figure}
\epsscale{1}
\plotone{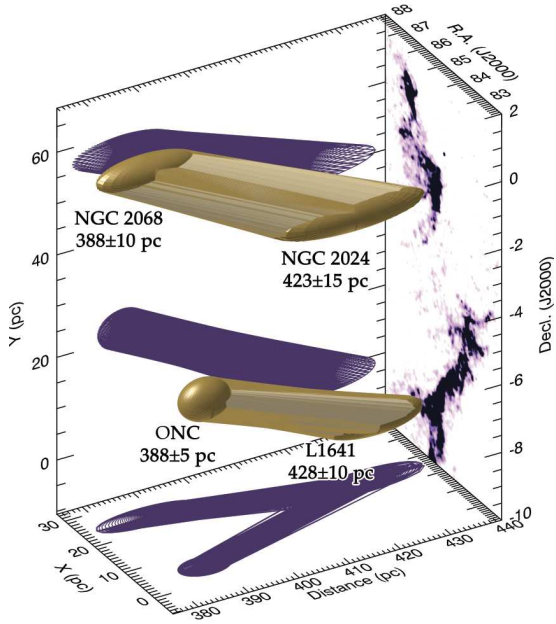} 
\caption{3d model of the Orion Complex. The width of the end ellipsoids in the model along the distance is representative of the uncertainties in the measurement and not the actual depth of each cluster. The plane of the sky plane shows the
extinction map from \citet{2011gutermuth}. Purple shadows are the projections of the model onto the remaining planes. Conversion of the length along the plane of the sky to pc is done at the distance of 388 pc.\label{fig:map}}
\end{figure}

The weighted average distance with the weighted uncertainty of all the stars located towards the ONC, including the Trapezium, is 388$\pm$5 pc. The distance measurements to nearly all stars is consistent with the average distance for the cluster (Figure \ref{fig:bar}). There is some scatter in the individual measurements of distance. Most of this scatter is likely to be systematic in nature. Some of this scatter may be physical, as the stars detected towards the ONC span a 4 pc region in the plane of the sky at the distance of 388 pc. It is also possible that some substructure is present in the ONC, however, due to limited sampling this possibility is not definitive.

The southern end of L1641 appears to be located considerably farther away, at 428$\pm$10 pc (Figure \ref{fig:map}). We include the effect of the dispersive delay in this value, which we estimate to be comparable to what we found in the ONC. Unfortunately we cannot measure it directly, as all of the secondary calibrators are located too far away from the targeted YSOs for their positions to be useful.

The exact manner in which L1641 connects to the ONC is unclear as there are no galactic sources detected in the northern part of the filament. However, it is not unreasonable to assume that the northern part of L1641 filament should be located at a similar distance to the ONC. There is a smooth gradient in radial velocity (RV) along the Orion A molecular cloud, ranging from $\sim$8 to 4 km s$^{-1}$ from northern to southern end of the L1641 \citep{1987bally, 2015nishimura}. This could imply either a passive rotation of the cloud, moving from very inclined to a more face-on orientation, or it could be the result of something actively pushing on the gas causing it to accelerate \citep[it is notable that the Orion Complex lies near the edge of the Orion-Eridanus superbubble,][]{2015Ochsendorf}. If the latter is true, it could potentially explain a large number of stars in the ONC that appear to be blue-shifted relative to the molecular gas \citep{2016kounkela}, as their RV would be representative of the initial velocity of the gas rather than the current velocity.

It is difficult to determine how accurate our measurement of the distance towards NGC 2068 is, as all of the calibrators, including the primary calibrator, are over 3$^\circ$ away. We estimate the effect of the dispersive delay to be on the order of 0.066 mas or 10 pc at the distance of NGC 2068. We find a distance to NGC 2068 of 388$\pm$10 pc. Finally, a distance towards NGC 2024 at this time cannot be reliably measured given the high incidence of multiplicity, as well as a limited number of observations currently available for the stars found towards this region. Neither can we currently obtain a reliable distance towards the $\sigma$ Ori cluster. This would be resolved with further monitoring. At this time we estimate the distance towards NGC 2024 on the basis of measurements towards VLBA 61/62 and VLBA 148 to be 423$\pm$15 pc.

\subsection{Proper motions and runaway stars}\label{subsec:pm}

\begin{figure}
\epsscale{1}
\plotone{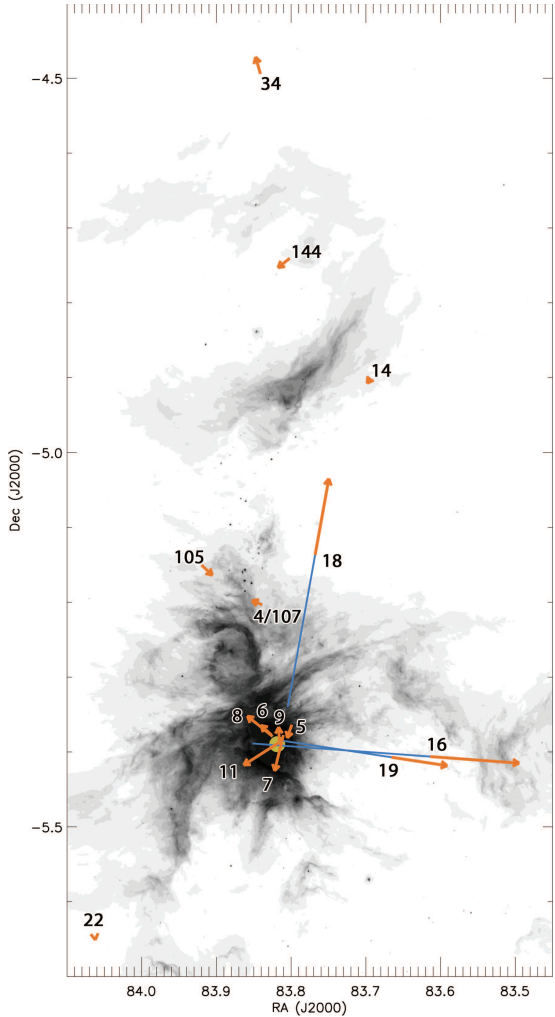} 
\caption{Proper motion vectors of the stars detected towards the ONC, corrected for the average motion of the cluster ($\mu_\alpha=1.38$ mas year$^{-1}$, $\mu_\delta=-0.36$ mas year$^{-1}$). The length of the vectors corresponds to motions over $5\times10^4$ years. The yellow dot at the center shows the current position of $\theta^1$ Ori C. Blue lines show the trajectory of the runaway stars over the last $10^5$ years. All the sources are labeled with their VLBA number. The greyscale background is the 8 $\mu$m Spitzer map from \citet{2012Megeath}.\label{fig:pmonc}}
\end{figure}

\begin{figure}
\epsscale{1}
\plotone{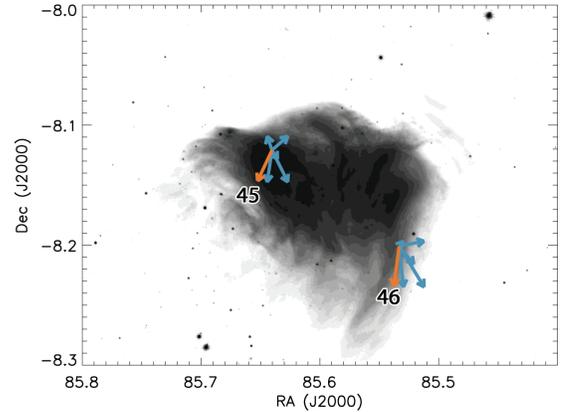} 
\caption{Proper motion vectors of the stars detected towards L1641 in the local standard of rest reference frame. The length of the vectors corresponds to motions over $5\times10^4$ years. Orange vectors are the measured proper motions, blue vectors are motions relative to the average motion of the ONC with a combination of $\pm$1$\sigma$ formal uncertainty of the average motion of the ONC in both $\mu_\alpha$ and $\mu_\delta$. All the sources are labeled with their VLBA number. The greyscale background is 8 $\mu$m Spitzer map from \citet{2012Megeath}\label{fig:pml1641}}
\end{figure}

\begin{figure}
\epsscale{1}
\plotone{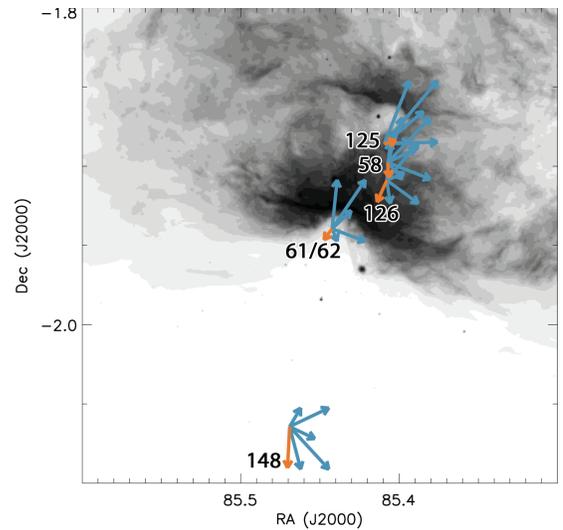} 
\caption{Same as Figure \ref{fig:pml1641}, but for NGC 2024.\label{fig:pmn2024}}
\end{figure}

\begin{figure}
\epsscale{1}
\plotone{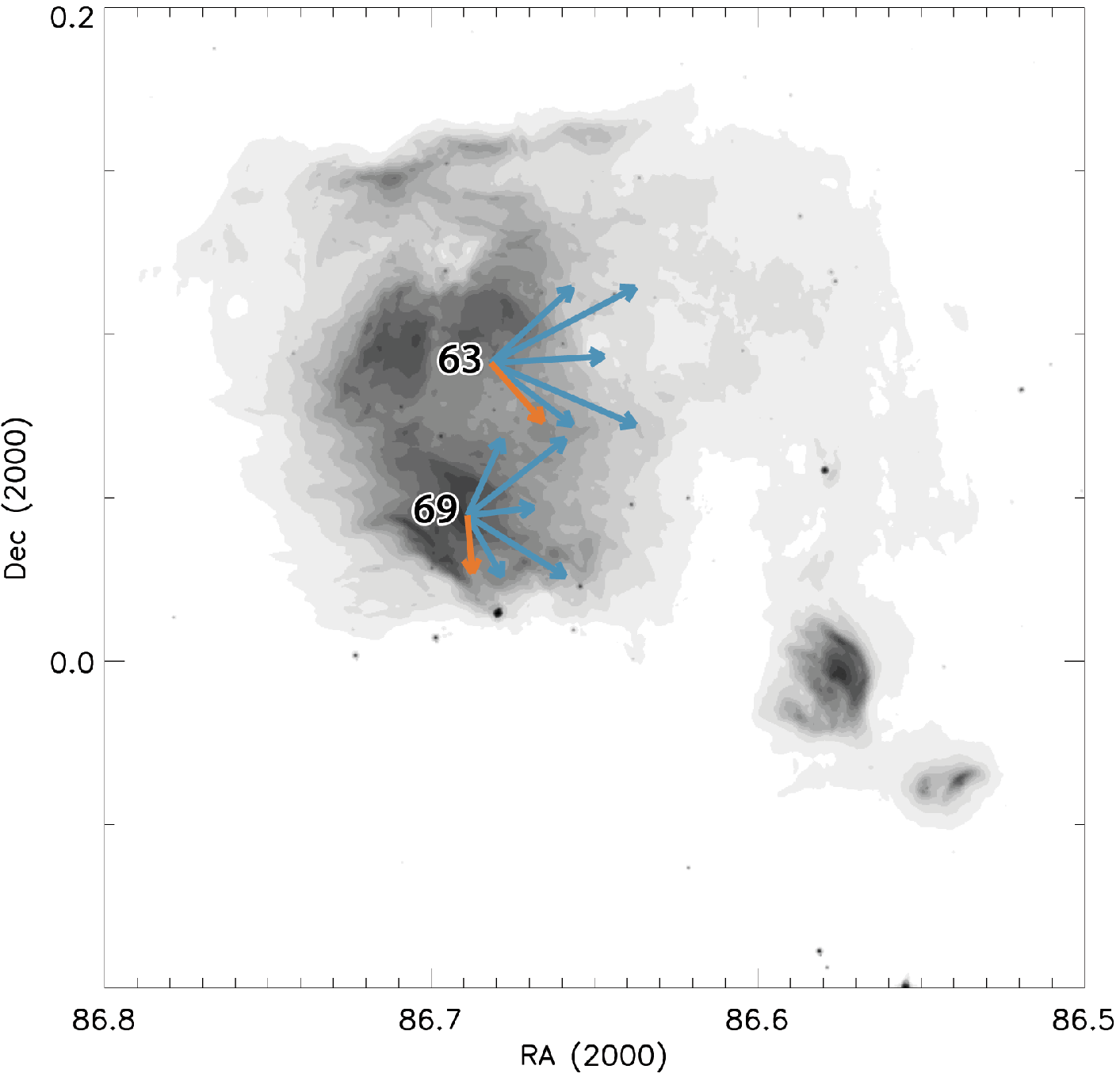} 
\caption{Same as Figure \ref{fig:pml1641}, but for NGC 2068.\label{fig:pmn2068}}
\end{figure}

Parallax and proper motion solutions ($v_{pm}$) are available for 15 systems in the ONC. While this is insufficient to perform a detailed analysis of the kinematics of the region, it is possible to obtain mean motions of the cluster.

Proper motions of long period binaries (i.e. VLBA 8, 9, 11, \& 4/107) are not representative of the $v_{pm}$ of the cluster, as they are significantly affected by the orbital motion. Therefore, we do not include them in the calculation of the mean. There are three stars that can be considered significant outliers in terms of proper motions. VLBA 16, 18, \& 19 have $v_{pm}$ range between 4.2 to 7.3 mas year$^{-1}$ (7.7 to 13.4 km s$^{-1}$). The difference is much larger than the typical dispersion velocity of 2.5 km s$^{-1}$ found towards the ONC \citep{2016kounkela}, therefore, they are also not included in the calculation of the proper motion of the cluster. The remaining 7 stars suggest $v_{pm}$ for the ONC of $\mu_\alpha^{lsr}=1.35\pm0.70$ mas year$^{-1}=2.49\pm1.29$ km s$^{-1}$ and $\mu_\delta^{lsr}=-1.44\pm1.51$ mas year$^{-1}$=$-2.66\pm2.79$ km s$^{-1}$, in the local standard of rest reference frame, corrected for the peculiar motion of the Sun (Figure \ref{fig:pmonc}). The uncertainties are obtained from the variance in the individual measurement, although they could be somewhat overestimated as the variance is largely driven by the peculiar velocity of stars within a cluster.

The most likely explanation for the high $v_{pm}$ for VLBA 16, 18, \& 19 is that they have been ejected from the cluster through a dynamical interaction within the cluster core, with the most notable suspect being $\theta^1$ Ori C. Unfortunately this star has not been detected by this program, but assuming that its proper motion should be similar to the average proper motion of the cluster, all three runaway stars appear to originate in the vicinity of it. Assuming linear motion, VLBA 16 appears to have been ejected $\sim 8\times10^4$ yr ago, VLBA 18 $\sim 12\times10^4$ yr ago, and VLBA 19 $\sim 8\times10^4$ yr ago. Some deceleration probably has occurred as they moved through the cluster; however, assuming the potential calculated by \citet{1998Hillenbrand}, this deceleration is not significant compared to the current $v_{pm}$ of these stars.

VLBA 16, 18, \& 19 are not alone in suffering the fate of being runaway stars. \citet{2005Poveda} identify JW 451 and 349 having $v_{pm}$ of 57 and 31 km s$^{-1}$, which also appear to originate from $\theta^1$ Ori C 1000 and 6000 yr. ago respectively. More famously, sources BN, I and n in the Orion BN/KL nebula have been accelerated to speeds of up to 26 km s$^{-1}$ through a dynamical interaction 500 yr. ago \citep{2008gomez,2011Goddi}. Even on a more extreme case, $\mu$ Col, AE Aur, and the compact binary $\iota$ Ori are thought to be ejected from the Trapezium cluster some 2.5 Myr ago through a four-body interaction \citep{2001deZeeuw, 2004Gualandris}. 

The average proper motion of L1641 is $\mu_\alpha^{lsr}=0.82\pm0.39$ mas year$^{-1}=1.67\pm0.79$ km s$^{-1}$ and $\mu_\delta^{lsr}=-2.20\pm0.38$ mas year$^{-1}$=$-4.48\pm0.78$ km s$^{-1}$. The southern end of the cloud appears to move westward relative to the ONC, although it is not collapsing into the ONC directly (Figure \ref{fig:pml1641}).

Analysis of the proper motions of NGC 2024 is once again made more complex by the multiplicity of the sources. The motions of VLBA 58, 125 \& 126, if they are indeed binaries, would be at least partially affected by the orbital motion. We obtained an orbital fit for VLBA 61/62, but the proper motions remain to be rather uncertain. Finally, VLBA 148 appears to be moving away from the cluster, and its motion appears to be rather distinct from however uncertain motions of the other stars. It is possible that it could have been ejected through dynamical interactions in the cluster, although further monitoring would be necessary to confirm it (Figure \ref{fig:pmn2024}).

The proper motion of NGC 2068 based on VLBA 63 and 69 is $\mu_\alpha^{lsr}=-0.62\pm0.73$ mas year$^{-1}=-1.15\pm1.34$ km s$^{-1}$ and $\mu_\delta^{lsr}=-1.27\pm0.02$ mas year$^{-1}=-2.35\pm0.04$ km s$^{-1}$ (Figure \ref{fig:pmn2068}). It presently appears to move towards NGC 2024, although this does not take into account the relative velocities of the two clusters.

\begin{figure}
\epsscale{0.7}
\plotone{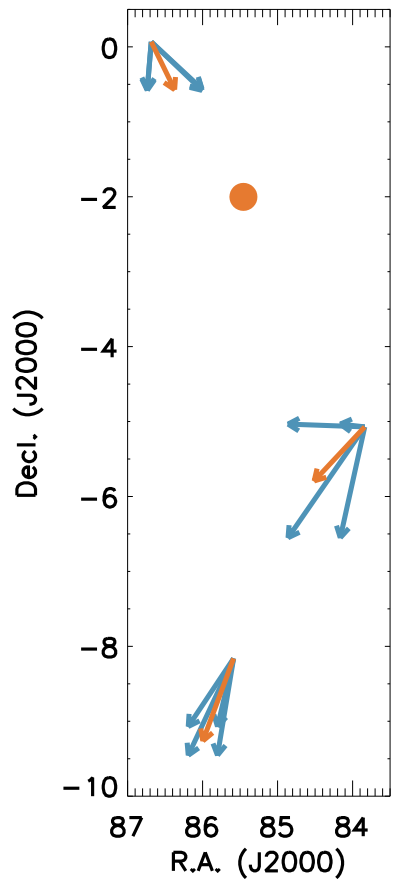} 
\caption{Orange arrows: lsr proper motions of the regions of the Orion Complex. Blue arrows: 1$\sigma$ uncertainty range in these values. Orange dot shows location of NGC 2024.\label{fig:pm}}
\end{figure}

It is clear that the entire complex appears to move in the southern direction on the equatorial globe, or in the direction of galactic rotation (Figure \ref{fig:pm}). Orion A also moves preferentially towards the East (towards the galactic plane), and the ONC has the largest eastward velocity compared to the other regions of the complex.

\section{Conclusions}

We monitored 36 non-thermal radio emitting YSOs spread throughout the Orion Complex with VLBA over a period of 2 years, and we report measured stellar parallaxes towards 26 of them. Fifteen of them are located towards the ONC, and we find a distance of 388$\pm$5 pc to the cluster; somewhat closer than the canonical $414\pm7$ pc distance found by \citet{2007menten} that is typically used in the literature. This result has implications on the luminosity and ages of the cluster. If the cluster is 7\% closer than the previous estimation, this implies that it is 12\% fainter, and 20\% older (assuming a relation $t\propto L^{-3/2}$) than what was previously reported in surveys of the ONC such as the one by \citet{2010dario}.

We also report distances towards other regions located in the Orion Complex, such as L1641, NGC 2024, and NGC 2068. While these values are somewhat more uncertain due to significantly smaller sample size, limited spatial coverage (particularly in case of L1641), and multiplicity, these are the first direct measurements of the stellar parallaxes towards these regions. This provides insight into the structure of the Complex.

We identify a possible region of large degree of plasma scattering towards the $\lambda$ Ori star forming region. The degree of scattering is significant, with broadening of the observed size of the objects of up to 16.5 mas near the center of the cluster at 5 GHz. The scattering is spread all within the ring 2.5---3$^\circ$ in radius produced by supernova activity. Unfortunately, this effect made it impossible to measure astrometry accurately enough to obtain a parallax towards stars found in this region.

A persistent problem in the analysis of both the parallax and proper motions of the stars is the multiplicity. We conclusively identify 5 of 27 stars that have been detected in at least three epochs belonging to a multiple system with orbital periods between 6 months to 10 years, with at least three more systems identified as likely binaries, although further monitoring would be necessary to confirm them. It is impossible to accurately determine parallaxes to these systems without solving for orbital motion of these systems, which we can presently do to only two of them. Six stars are known spectroscopic binaries with very short periods (one of them also has an aforementioned intermediate period companion); possibly a larger number of them could have very close companions that are yet to be identified, particularly since few surveys of spectroscopic binaries have been performed in the Orion Complex outside of the ONC. While understanding of their orbits is not detrimental for finding the parallax, it could still influence the solution somewhat. Finally, four stars in the sample have long period companions, although only one of them does not have a closer companion in a higher order multiple system. These wide companions should not affect the solution for the parallax, although they do affect proper motions. 

In all, at least 14 (possibly more) of 27 stars observed with VLBA belong to multiple systems. Whether this multiplicity fraction is consistent with that for the entire Complex is not yet known as it is presently difficult to identify companions with intermediate periods towards Orion due to its distance. Future generations of high resolution optical and IR telescopes would make it possible to identify the full extent of multiplicity towards this region.

Further monitoring of the identified YSOs would be beneficial as due to variability in radio, currently only a limited number of detections are available to some stars. In the future it would be possible to more effectively measure their parallax and proper motions. It will also be necessary to confirm multiplicity and constrain orbital parameters towards some sources.

The distance solutions produced by the GOBELINS survey will be used as independent constraint on the accuracy of \textit{Gaia}, as the systematic effects behind the sample selection and the individual observations are different between these two programs. Approximately half of the systems observed with VLBA towards the Orion Complex are optically visible, therefore it should be possible to compare the distance solutions towards them directly, at least in the ONC, although nebulosity could significantly degrade performance in the optical regime. On the other hand, star forming regions towards Orion B suffer from high extinction; therefore only a few members of NGC 2024 and NGC 2068 would be detectable with \textit{Gaia}.

\acknowledgments
We acknowledge helpful discussions with Mark Reid which led to testing the possible effects of phase delay in the ONC. G.N.O.-L., L.L., L.F.R., R.A.G.-L.L., G.P., and J.L.R. acknowledge DGAPA, UNAM, CONACyT, Mexico for financial support. L.L. and G.N.O.-L. also acknowledge support from von Humboldt Stiftung. N.J.E. was supported by NSF grant AST-1109116 to the University of Texas at Austin. P.A.B.G. acknowledges financial support from FAPESP. The National Radio Astronomy Observatory is operated by Associated Universities, Inc., under cooperative agreement with the National Science Foundation.

\input{obs.tex}
\input{calib.tex}
\input{pos.tex}
\input{par.tex}

\appendix
\section{$\lambda$ Ori scattering}

As with all the other regions, $\lambda$ Ori was observed at 5 GHz. During the first epoch of the observations J0536+0944 was used as a primary calibrator, which is located only 0.4$^\circ$ from the center of the cluster. Calibrating on this source could produce coherent signal only on the baselines shorter than 1000 km, that is the baselines between Fort Davis, Kitt Peak, Los Alamos, Pie Town, and partially Owens Valley.

The second and third epochs were calibrated on J0532+0732, located 2.4$^\circ$ from the cluster center, but still barely inside the $\lambda$ Ori ring. While this produced a significant improvement on the calibration over the first epoch, any baselines involving antennas at Hancock and St. Croix could not be calibrated. Since the longest baseline of VLBA was not used, uncertainties in source positions remain large. However, surprisingly, some baselines longer than the baselines involving these two antennas (including most baselines involving Mauna Kea) did produce some coherent signal.

To determine the cause of the poor signal, and potentially finding a primary calibrator that could cause an improvement on the data, in August 2015 we observed 4 calibrators, J0536+0944, J0532+0732, J0544+1118, and J0547+1223 at both 5 and 8 GHz. J0547+1223 was known to produce good calibration, it is located almost 4$^\circ$ from the cluster center. While it did appear as a point source at both wavelengths, such a large angular separation is larger than what is ideal for a primary calibrator. J0544+1118, located at 2.8$^\circ$ from the cluster center had an appearance very similar to J0532+0732 -- baselines involving HN and SC could not produce a coherent signal at 5 GHz. With the exception of J0547+1223, all 3 calibrators showed a significant improvement at 8 GHz, however, even at this wavelength they did not appear like point sources. And, even at 8 GHz, J0536+0944 did appear to be significantly poorer than any other calibrator. 

The Gaussian model fits of the sizes of these sources is listed in Table \ref{tab:laori}, these sizes are roughly consistent with $\lambda^2$. Because of this wavelength dependence, we believe that all radio observations of $\lambda$ Ori are significantly affected by the plasma scattering. A possible source of this scattering is the ionized gas in the $\lambda$ Ori ring that is currently 2.5---3$^\circ$ in radius (Figure \ref{fig:all}), left behind by a supernova blast that originated 1 Myr ago \citep{2002Dolan}. However, it is surprising that $\lambda$ Ori is the only Orion region where scattering is a concern. Very few regions are known to be sources of significant plasma scattering processes that can be observed in this portion of radio regime; among them are the Galactic Center \citep{1998Bower}, NGC 6334 \citep{2012rodriguez}, and Cygnus \citep{2001Desai}.

The fourth epoch of the observations was done at 8 GHz and it used J0544+1118 as the primary calibrator. Only 2 sources were detectable at a higher frequency (VLBA 85 and 87). After considering positional offsets that were introduced with the several alterations of the primary calibrators, neither of these sources appear to be galactic. No further monitoring was done for the region.

One object of interest identified towards $\lambda$ Ori is VLBA 85. It has two components separated by 0.7 mas, possible AGN jet.
\input{laori.tex}

\section{Comments on the individual sources in the ONC}
\subsection{Trapezium}

\textbf{VLBA 5} (=GMR A) is found to be at a distance of 374$\pm$15 pc with a proper motion of $\mu_\alpha=1.81\pm 0.11$ mas year$^{-1}$ and $\mu_\delta=-1.62\pm 0.13$ mas year$^{-1}$. It has also been previously monitored with VLBA in a period from Jan. 2003 to Dec. 2004 by S07 and from Sept. 2006 to Mar. 2007 by MR. S07 found a distance solution to GMR A of 389$^{+24}_{-21}$ pc with $\mu_\alpha=1.89\pm 0.12$ mas year$^{-1}$ and $\mu_\delta=-1.67\pm 0.19$ mas year$^{-1}$, which is consistent with the distance found in this work. On the other hand, MR found a distance solution of 418.4$\pm$18.2 pc with $\mu_\alpha=1.82\pm 0.09$ mas year$^{-1}$ and $\mu_\delta=-2.05\pm 0.18$ mas year$^{-1}$, a discrepancy on the order of 2$\sigma$ in distance and $\mu_\delta$. The difference in proper motion is unlikely to be attributed to a long-period multiplicity due to the lack of acceleration observed between S07 and this work. No information is available in regards to whether or not GMR A belongs to a compact binary.

\textbf{VLBA 6} (=GMR F) is observed to be at a distance of 389$\pm$7 pc with a proper motion of $\mu_\alpha=2.38\pm 0.08$ mas year$^{-1}$ and $\mu_\delta=0.55\pm 0.14$ mas year$^{-1}$. MR find a distance solution of 406.1$\pm$8.4 pc with $\mu_\alpha=2.24\pm 0.09$ mas year$^{-1}$ and $\mu_\delta=-0.66\pm 0.18$ mas year$^{-1}$. While the two distance estimates disagree, the proper motions are consistent, suggesting that GMR F is unlikely to be a long-period binary. It has been identified as a double-lined spectroscopic binary by \citet{2002Prato} with $q\sim0.31$, but no orbital solution is available.

\textbf{VLBA 7} (=GMR H) was detected only in the first three epochs of the observations; therefore uncertainties in the solution are presented based only on the astrometric uncertainties and do not take into the account the systematic offsets. We obtain a distance solution of 372$\pm$15 pc and $\mu_\alpha=2.22\pm 0.18$ mas year$^{-1}$ and $\mu_\delta=-3.08\pm 0.55$. Not observed by MR.

\textbf{VLBA 8} (=GMR G) is found to be at a distance of 377$\pm$5 pc with $\mu_\alpha=3.82\pm 0.10$ mas year$^{-1}$ and $\mu_\delta=1.60\pm 0.17$ mas year$^{-1}$. MR only observed this object in three epochs. For this reason they do not present a distance solution; however, the data taken by their program suggest a distance of 382$\pm$4 pc, providing a good agreement to the distance obtained through in this work. On the other hand, proper motions obtained by MR are $\mu_\alpha=4.29\pm 0.17$ mas year$^{-1}$ and $\mu_\delta=3.33\pm 0.37$ mas year$^{-1}$. This shows that this star underwent a significant acceleration in the 8 years between these observations, suggestive of a long-period binary; however, the magnitude of the acceleration is not sufficient to noticeably affect the proper motions during the $\sim$ 2 years covered by either program separately. No optical or IR companion to the system has been previously identified.

\textbf{VLBA 9} (=GMR 25, = $\theta^1$ Ori E) is a known spectroscopic binary \citep{2008Costero,2012Morales} with a circular orbit, period of 9.89520 days, $q\sim1$, $i=74^\circ$ and masses of 2.807 and 2.797 M$_\odot$. Not previously observed by MR. We find a distance solution of 391$\pm$ 7 pc with $\mu_\alpha=1.45\pm 0.03$ mas year$^{-1}$ and $\mu_\delta=1.02\pm 0.08$ mas year$^{-1}$, without taking into the account the orbital motion. The lack of eccentricity makes it difficult to constrain the position of the star in its orbit with the astrometric data, but the typical effect of this orbit on the distance solution is within 3 pc.

\textbf{VLBA 11} (=GMR 12, = $\theta^1$ Ori A) is a known triple system. One of the components has been detected through the adaptive optics and VLTI imaging \citep{2013Close,2013Grellmann}, with the projected separation of $\sim0.2''$ or $\sim$70 AU. The other component is an eclipsing system that is also observed through spectroscopy with a period of 65.4 days \citep{2000Stickland}. Not accounting for the orbital motion we obtain a distance solution of 380$\pm$7 pc with $\mu_\alpha=4.81\pm 0.07$ mas year$^{-1}$ and $\mu_\delta=-2.33\pm 0.09$ mas year$^{-1}$. Incorporating the orbital motion of the spectroscopic binary while solving for $i$ and $\Omega$ yields comparable distance of 380$\pm$14 pc with $\mu_\alpha=4.81\pm 0.10$ mas year$^{-1}$, $\mu_\delta=-2.53\pm 0.12$, $\Omega=150\pm20^\circ$ and $i=87\pm11^\circ
$, therefore, consistent with an eclipsing system. This does not change significantly by assuming that the emission is coming from the secondary instead of the primary. MR found a distance towards GMR 12 of 417.9$\pm$9.2 pc with $\mu_\alpha=4.82\pm 0.09$ mas year$^{-1}$ and $\mu_\delta=-1.54\pm 0.18$ mas year$^{-1}$. This solution did not take into the account the orbital motion; however, including it does not significantly alter the fit. The distances between two observations are discrepant by $\sim2\sigma$. The difference in the measured proper motion is most likely driven by the acceleration due to the orbital motion of the long-period binary.

Other confirmed galactic sources that have been detected towards the Trapezium include VLBA 13, 149, and 150, but as they have been detected in only two epochs, currently it is impossible to do a parallactic fit.

\subsection{Outside of the Trapezium}

\textbf{VLBA 4/107} (=Brun 656) is located westward of the OMC 2/3 filament. It was detected as an astrometric binary system, with VLBA 4 detected in epochs 1, 4, \& 5, and VLBA 107 detected in epochs 2, 4, \& 5. Orbital motion is clearly apparent in both stars. However, with only six positions it is impossible to fit all 13 parameters for both parallactic and orbital motion. Therefore we exclude $i$ and $\Omega$ from the fit, assuming a face on orientation. In the follow up work with additional data it would be possible to present a full solution. Potentially a member of a triple system as \citet{2006kohler} identify an additional companion 0.4'' or $\sim$150 AU away; this should have little impact on astrometry after 2 year baseline. We obtain a distance solution of 369$\pm$10 pc with the proper motions of the compact system of $\mu_\alpha=2.36\pm 0.69$ mas year$^{-1}$ and $\mu_\delta=0.06\pm 1.05$ mas year$^{-1}$, and the scatter and the dependence between these parameters is shown in the Figure \ref{fig:bin}. This proper motion is most likely not representative of the true proper motion of the triple system. We find the period of the compact system to be 6.27$\pm$0.54 years and $M\cos^3 i$ of both components of 1.70$\pm$0.16 and 1.62$\pm$0.38 M$_\odot$. The spectral type for the primary has been previously reported to be G2III \citep{1997hillenbrand}

\textbf{VLBA 14} (=V1699 Ori) is located towards NGC 1977. We found a distance solution of 401$\pm$7 pc with $\mu_\alpha=1.76\pm 0.05$ mas year$^{-1}$ and $\mu_\delta=-0.89\pm 0.16$ mas year$^{-1}$.

\textbf{VLBA 16} (=Parenago 1469) is located westward of the Trapezium. It was detected only in the epochs 1, 2, \& 4. We find a distance solution of 394$\pm$4 pc with $\mu_\alpha=-7.22\pm 0.06$ mas year$^{-1}$ and $\mu_\delta=-0.99\pm 0.08$ mas year$^{-1}$. Proper motions for this star are uncommonly large and appear to be projected from the Trapezium cluster (see Section \ref{subsec:pm}).

\textbf{VLBA 18} (=Parenago 1724) is located westward of the OMC 2/3 filament. We found a distance solution of 398$\pm$7 pc with $\mu_\alpha=0.06\pm 0.20$ mas year$^{-1}$ and $\mu_\delta=6.95\pm 0.15$ mas year$^{-1}$. Similarly to VLBA 16, it also has a very large proper motion vector, which projects back to the center of the Trapezium cluster. It was previously identified by \citet{1998Neuhaeuser} as a runaway star.

\textbf{VLBA 19} (=Parenago 1540) is located westward of the Trapezium. It is a known double-lined spectroscopic binary \citep{1988Marschall}. Without accounting for the orbital motion we obtain a distance solution of 404$\pm$11 pc with $\mu_\alpha=-3.88\pm 0.13$ mas year$^{-1}$ and $\mu_\delta=-1.10\pm 0.15$ mas year$^{-1}$. Incorporating an orbital fit and solving for $i$ and $\Omega$ yields a distance of 386$\pm$7 pc with $\mu_\alpha=-4.01\pm 0.08$ mas year$^{-1}$, $\mu_\delta=-1.17\pm 0.07$ mas year$^{-1}$, $i=104\pm12^\circ$ and $\Omega=69\pm11^\circ$. This results in the masses of the components of 0.49$\pm$0.10 and 0.37$\pm$0.07 M$_\odot$, however, the spectral types of K3V and K5V make these masses to be somewhat suspect. With an assumption that the emission is coming from the secondary with $\omega$ rotated by 180$^\circ$, an alternate family of solutions is found at 413$\pm$12 pc with $\mu_\alpha=-3.85\pm 0.14$ mas year$^{-1}$, $\mu_\delta=-1.12\pm 0.08$ mas year$^{-1}$, $i=56\pm12^\circ$ and $\Omega=130\pm14^\circ$, and masses of 0.78$\pm$0.40 and 0.60$\pm$0.28 M$_\odot$. Similarly to VLBA 16 and 18, this system has a very large proper motion vector that projects back to the Trapezium cluster, and it has been previously theorized to be a runaway star by \citet{1988Marschall}.

\textbf{VLBA 22} (=HD 37150) is located towards the south-east of the Trapezium. We obtain a distance solution of 394$\pm$7 pc with $\mu_\alpha=1.32\pm 0.05$ mas year$^{-1}$, $\mu_\delta=0.56\pm 0.12$ mas year$^{-1}$.

\textbf{VLBA 27/28} (=NU Ori) shows considerable motion, but it cannot be fitted yet due to multiplicity. VLBA 27 has been detected in epochs 1, 3, \& 5, and VLBA 28 has been detected in epoch 1, with the projected separation of 35 mas from VLBA 27. A single point source was detected in epoch 4, tentatively interpreted to be associated with VLBA 28, $\sim$8 mas away from the expected position of VLBA 27. This companion system has been previously predicted to exist by \cite{2013Grellmann} based on the VLTI observations. In addition to this, this system contains a known spectroscopic binary with a period of 19 days and $a_1\sin i$ of $\sim 0.15$ mas \citep{1991Abt}; a hint of extension is seen towards VLBA 27 in some epochs and this could be the source. This system also has a wider companion with a separation of $\sim 0.5''$ \citep{2006kohler}. Given the fact that the primary was detected only in spring epochs without any fall epochs, we cannot provide even rough constraints on its distance. Follow-up monitoring would be needed in order to accurately understand the motions of this system.

\textbf{VLBA 34} (=HD 37017) is a known double-lined spectroscopic binary \citep{1998Bolton}. Without orbital motion we obtain a distance solution of 360$\pm$7 pc with $\mu_\alpha=1.87\pm 0.07$ mas year$^{-1}$, $\mu_\delta=1.17\pm 0.24$ mas year$^{-1}$. Solving for $i$ and $\Omega$ we obtain two possible results due to the lack of constrains in the direction of the orbit. These results are $i=53\pm23^\circ$ and $i=127\pm28^\circ$, both with $\Omega=131\pm26^\circ$ and the distance solution of 378$\pm$10 pc with $\mu_\alpha=1.88\pm 0.09$ mas year$^{-1}$, $\mu_\delta=1.20\pm 0.14$ mas year$^{-1}$. This corresponds to masses of the components of 4.09$\pm$4.41 and 2.14$\pm$2.22 M$_\odot$. Assuming that the emission is coming from the secondary does not change the solution significantly, the estimated distance becomes 383$\pm$5 pc, although the inclination angle becomes $76\pm18^\circ$.

\textbf{VLBA 105} (=	Parenago 2148) is located towards OMC 2/3 filament. It was detected only in epochs 2, 3, \& 5. The positional uncertainties, particularly in epoch 2, are rather substantial as the source appears to be marginally extended in $\alpha$. Without accounting for any of the systematic offsets we find a distance solution of 388$\pm$53 pc with $\mu_\alpha=0.33\pm 0.05$ mas year$^{-1}$, $\mu_\delta=-1.34\pm 0.43$ mas year$^{-1}$. 

\textbf{VLBA 114} (=	Parenago 1778) is located towards NGC 1977. It was not detected in epoch 1. We obtain a distance solution of 437$\pm$83 pc with $\mu_\alpha=2.54\pm 0.30$ mas year$^{-1}$, $\mu_\delta=-1.30\pm 0.64$ mas year$^{-1}$. The fit is rather poor, despite the substantial positional uncertainties. This could be attributed to acceleration due to multiplicity.

Other galactic sources identified in the region that have been detected only in two epochs are VLBA 33, 110, \& 116. VLBA 115 has also been detected in 2 epochs, and although the $\alpha$ offset is consistent with belonging to the ONC (2.7 mas), the $\delta$ offset is over 13 mas; this is likely due to multiplicity.

\section{Comments on the remaining regions}
\subsection{L1641}

Only two galactic sources, VLBA 45 \& 46, have been detected towards L1641. Both of them are located on the southern end of the cloud, at $\delta<-8^\circ$. The solutions towards them are somewhat more uncertain than they are towards sources located within the ONC, with very large uncertainties and imprecise fit. Preliminary fits result in distance solutions of 424$\pm$12 pc with $\mu_\alpha=0.68\pm 0.06$ mas year$^{-1}$, $\mu_\delta=-0.31\pm 0.22$ mas year$^{-1}$ for VLBA 45 and 433$\pm$28 pc with $\mu_\alpha=0.25\pm 0.04$ mas year$^{-1}$, $\mu_\delta=-0.47\pm 0.25$ mas year$^{-1}$ for VLBA 46. The fit appears to be somewhat dubious and the observed positions do not agree with the best fits for both sources in the epochs 1 \& 4. It is possible that this offset is attributable to multiplicity in both of these objects, although the fact that the magnitude of the offset is comparable for both sources at each epoch makes it more dubious. Therefore we treat this offset as the pointing error due to calibration and solve for a common offset for both sources.

We correct the positions of the first epoch by $\Delta\alpha=0.256$ mas and $\Delta\delta=0.771$ mas, and the fourth epoch by $\Delta\alpha=0.204$ mas and $\Delta\delta=0.659$ mas. It is somewhat curious that this offset is comparable in both epochs. After this correction, the distance solution becomes 425$\pm$12 pc with $\mu_\alpha=0.68\pm 0.09$ mas year$^{-1}$, $\mu_\delta=-0.51\pm 0.25$ mas year$^{-1}$ for VLBA 45 and 431$\pm$13 pc with $\mu_\alpha=0.13\pm 0.25$ mas year$^{-1}$, $\mu_\delta=-1.05\pm 0.18$ mas year$^{-1}$ for VLBA 46. This offset has a very small overall effect on the distance, but the proper motions are somewhat uncertain.

In addition to the galactic sources, there are a number of extragalactic sources that can be of interest. VLBA 39/40, 41/42, 47/48, 89/90,  \& 94/95/96 appear to exhibit no motion between epochs, but they appear to be extended, double or even triple objects. They can probably be attributable to the AGN jets. It is curious that so many of these extended sources appear to be in the direction of this particular region.

\subsection{NGC 2068}

The second epoch of the observations of this region was strongly affected by a pointing error, and this offset is also present in nearby sources that can otherwise be considered extragalactic. A possible explanation for this is that one of the secondary calibrators had an extremely weak detection in this epoch. We solve for a common offset for all sources of $\Delta\alpha=-0.559$ mas, $\Delta\delta=-0.515$ mas.

Three galactic sources have been detected towards NGC 2068. VLBA 63, 68 (=HD 290862), \& 69. After removing the offset, we obtain a distance solution of 383$\pm$7 pc with $\mu_\alpha=-1.02\pm 0.10$ mas year$^{-1}$, $\mu_\delta=-0.52\pm 0.15$ mas year$^{-1}$ for VLBA 63, and 392$\pm$5 pc with $\mu_\alpha=-0.01\pm 0.10$ mas year$^{-1}$, $\mu_\delta=-0.49\pm 0.08$ mas year$^{-1}$ for VLBA 69. On the other hand, we can offer only an extremely noisy tentative solution for VLBA 68 of 455$\pm$113 pc with $\mu_\alpha=0.35\pm 0.27$ mas year$^{-1}$, $\mu_\delta=0.83\pm 0.83$ mas year$^{-1}$. The reason for this is that VLBA 68 appears to be a multiple system (although no companion has been directly detected) which greatly affects the positions. At this point in time we cannot perform an orbital fit for the system.

\subsection{NGC 2024}
VLBA 61/62 has been identified as an astrometric binary, with the first component present in all five epochs, while the second component detected only in epochs 1, 4, \& 5. We find a distance to the system of 434$\pm$10 pc with $\mu_\alpha=0.47\pm 0.32$ mas year$^{-1}$, $\mu_\delta=0.39\pm 0.62$ mas year$^{-1}$, and the scatter and the dependence between these parameters is shown in the Figure \ref{fig:bin}. The period of the binary is 9.50$\pm$0.67 years, inclination of 141$\pm6^\circ$, and masses of both components of 1.85$\pm$0.58 and 0.95$\pm$0.22 M$_\odot$. Unfortunately, this system has not been detected at any other wavelength regime other than in radio, therefore, a comparison of masses to spectral types is impossible. While the orbital fit itself is convergent, it must be noted that some uncertainty does remain due to the limited number of detections, and although unlikely, a possibility of a somewhat larger distance $\sim$444 pc as well as a somewhat steeper inclination angle cannot be ruled out currently.

VLBA 58 is another star that was detected in all five epochs in this region, and it also appears to be a binary due to its peculiar motions from one epoch to the next. At this point in time, an orbit to it cannot be fitted, but we obtain a very tentative solution of 449$\pm$24 pc with $\mu_\alpha=0.04\pm 0.31$ mas year$^{-1}$, $\mu_\delta=0.20\pm 0.43$ mas year$^{-1}$.

Three other stellar objects have been detected towards the region, but only in three epochs, therefore, solutions are somewhat uncertain as they do not take into account any systematic offsets. VLBA 148 has a distance solution of 412$\pm$ 5 pc with $\mu_\alpha=0.19\pm 0.44$ mas year$^{-1}$, $\mu_\delta=-0.97\pm 0.27$ mas year$^{-1}$. VLBA 125 is presently found at a distance of 536$\pm$30 with $\mu_\alpha=-0.43\pm 0.16$ mas year$^{-1}$, $\mu_\delta=1.03\pm 0.42$ mas year$^{-1}$; this solution is rather tentative due to the astrometric errors and a quality of the fit. On the other hand, VLBA 126 has a measured distance of 356$\pm$4 pc with $\mu_\alpha=0.55\pm 0.10$ mas year$^{-1}$, $\mu_\delta=-0.10\pm 0.15$ mas year$^{-1}$.

It is clear that VLBA 125 and 126 have a measured distance that is decidedly different from what is found towards other objects in the region. A possible explanation to this is that these stars may belong to an as of yet unseen binary system. Whether this is could also be the case for VLBA 148 is as of yet unclear.

Other galactic sources identified towards NGC 2024 are VLBA 124 \& 153, although only two epochs are currently available.

A number of extragalactic objects of interest have also been identified. VLBA 56 shows a clear signature of an AGN jet. VLBA 146/147 has two components, also a probable extragalactic jet. While previously VLBA 55 has been identified as a Class II YSO based on its infrared colors, it shows no positional offset between epochs. It was notable for being extremely bright in radio (highest VLBA flux of 338 mJy at 5 GHz).

\subsection{$\sigma$ Ori}

Only one Galactic object has been identified towards $\sigma$ Ori among those detected with VLBA. VLBA 145 (=HD 294300) was not monitored in epochs 1 \& 2, therefore currently only three epochs of astrometry are available. It is found at a distance of 302$\pm$32 pc with $\mu_\alpha=-4.92\pm 0.66$ mas year$^{-1}$, $\mu_\delta=4.67\pm 1.37$ mas year$^{-1}$. Whether this distance solution is trustworthy or not still remains to be tested; while $\sigma$ Ori is most likely spatially separate from NGC 2024, a difference in distance of over 100 pc would be surprising. \citet{2008Sherry} previously estimated distance towards $\sigma$ Ori based on main-sequence fitting to be 420$\pm$30 pc, quite close to the distances we obtain to NGC 2024 members. Combined with the rather high proper motions for VLBA 145 of 9.7 km s$^{-1}$ as well as a somewhat poor fit could imply that this star belongs to a binary system; therefore, further monitoring would be needed to better understand the kinematics of the system.

\subsection{L1622}

In the VLA survey, only two sources have been identified towards L1622, of which one was a known YSO, and one did not have any classification. The former one was not detected with VLBA, the latter (VLBA 84) did not exhibit any positional offset between observations. No parallax towards this region can be measured. No further monitoring was done past epoch 2.

\bibliographystyle{apj.bst}


\end{document}

%% file: obs.tex
\begin{deluxetable*}{ccccccccc}
\tabletypesize{\scriptsize}
\tablewidth{0pt}
\tablecaption{Dates of the VLBA observations. \label{tab:obs}}
\tablehead{
\colhead{\#} &\colhead{R.A.} &\colhead{Dec.} & \colhead{Epoch 1} & \colhead{Epoch 2} & \colhead{Epoch 3} & \colhead{Epoch 4} & \colhead{Epoch 5}\\
\colhead{} &\colhead{(J2000)} &\colhead{(J2000)} & \colhead{Date / Field name} & \colhead{Date / Field name} & \colhead{Date / Field name} & \colhead{Date / Field name} & \colhead{Date / Field name}
}
\startdata
1 & 05:35:13.5550 & -05:21:37.8875 & 03/24/14 GD1 & 09/21/14 EG1 & 02/26/15 FB1 & 09/01/15 FP1 & 02/25/16 FA1 \\
2 & 05:35:28.5199 & -05:10:11.9953 & 03/24/14 GD2 & 09/21/14 EG2 & 02/26/15 FB2 & 09/01/15 FP2 & 02/25/16 FA2 \\
3 & 05:35:57.8954 & -05:23:01.0259 & 03/24/14 GD3 & 09/21/14 EG3 & --- & --- & --- \\
4 & 05:35:32.0330 & -05:39:38.5578 & --- & --- &  02/26/15 FB3 & 09/01/15 FP3 & --- \\
5 & 05:34:51.3847 & -04:55:18.2572 & 03/10/14 GE1 & 09/22/14 EH1 & 02/27/15 FC2 & 08/31/15 FQ2 & 03/01/16 HV2 \\
6 & 05:34:51.3847 & -05:09:09.6415 & 03/10/14 GE2 & 09/22/14 EH2 & 02/27/15 FC3 & 08/31/15 FQ3 & 03/01/16 HV3 \\
7 & 05:34:31.9858 & -05:28:33.5795 & 03/10/14 GE3 & 09/22/14 EH3 & 02/27/15 FC1\tablenotemark{a} & 08/31/15 FQ1 & 03/01/16 HV1 \\
8 & 05:35:49.5816 & -05:16:05.3335 & 03/08/14 GF1 & 09/30/14 EI1 & 02/28/15 FD2 & 09/05/15 FR2 & 03/05/16 HW2 \\
\enddata
\tablenotetext{a}{Field was repositioned to RA=05:34:34.7570  DEC=-05:25:47.3027}
\end{deluxetable*}

%% file: calib.tex
\begin{deluxetable*}{ccc}
\tabletypesize{\scriptsize}
\tablewidth{0pt}
\tablecaption{Assumed positions of the calibrators\tablenotemark{a}\label{tab:poscalib}}
\tablehead{
\colhead{Calibrator} &\colhead{$\alpha$} &\colhead{$\delta$} \\
\colhead{} &\colhead{(J2000)} &\colhead{(J2000)}
}
\startdata
J0539-0514 & 05:39:59.937192 &  -05:14:41.30174  \\
J0529-0519 & 05:29:53.533450 &  -05:19:41.61678  \\
J0541-0541 & 05:41:38.083371 &  -05:41:49.42843  \\
J0532-0307 & 05:32:07.519261 &  -03:07:07.03799  \\
J0517-0520 & 05:17:28.110157 &  -05:20:40.84222  \\
J0542-0913 & 05:42:55.877408 &  -09:13:31.00660  \\
J0553-0840 & 05:53:41.891558 &  -08:40:01.90151  \\
J0527-1002 & 05:27:24.060380 &  -10:02:57.26651  \\
J0558-0055 & 05:58:44.391460 &  -00:55:06.92375  \\
J0600-0005 & 06:00:03.503368 &  -00:05:59.03477  \\
J0552+0313 & 05:52:50.101499 &  +03:13:27.24311  \\
J0536+0944 & 05:36:31.978172 &  +09:44:20.63128  \\
J0532+0732 & 05:32:38.998486 &  +07:32:43.34572  \\
J0519+0848 & 05:19:10.811128 &  +08:48:56.73450  \\
J0544+1118 & 05:44:52.199795 &  +11:18:49.92568  \\
J0530+1331 & 05:30:56.416747 &  +13:31:55.14954  \\
J0547+1223 & 05:47:06.276323 &  +12:23:46.24477
\enddata
\tablenotetext{a}{When imaged, secondary calibrators are offset from these positions as their coordinates are referenced to the primary calibrator.}
\end{deluxetable*}

\begin{deluxetable*}{ccccc}
\tabletypesize{\scriptsize}
\tablewidth{0pt}
\tablecaption{Calibrators setup for the observed fields\label{tab:calib}}
\tablehead{
\colhead{Field} &\colhead{Primary} &\colhead{Secondary} & \colhead{Secondary} & \colhead{Secondary} \\
\colhead{\#} &\colhead{} &\colhead{1} & \colhead{2} & \colhead{3}
}
\startdata
1, 2, 3, 4, 8, 9, 10, 11, 12, 13, 14, 15, 16, 17 & J0539-0514 & J0529-0519 & J0541-0541 & J0532-0307 \\
5, 6, 7, 35, 36 & J0529-0519 & J0539-0514 & J0532-0307 & J0517-0520 \\
18, 19, 20, 37 & J0539-0514 & J0541-0541 & J0529-0519 & J0542-0913 \\
21, 22, 23 & J0542-0913 & J0541-0541 & J0553-0840 & J0527-1002 \\
24, 25, 26 & J0532-0307 & J0539-0514 & J0541-0541 & J0558-0055 \\
27, 28, 29, 45, 46 & J0558-0055 & J0600-0005 & J0552+0313 & J0532-0307 \\
30, 31, 32, 33, 34 & J0532-0307 & J0539-0514 & J0529-0519 & J0558-0055 \\
38, 39 & J0536+0944 & J0532+0732 & J0544+1118 & J0530+1331 \\
40 & J0552+0313 & J0600-0005 & J0558-0055 & J0606-0024 \\
41 & J0552+0313 & J0532-0307 & J0558-0055 & J0606-0024 \\
42, 43, 44& J0541-0541 & J0539-0514 & J0529-0519 & J0542-0913
\enddata
\end{deluxetable*}

%% file: pos.tex
\begin{deluxetable*}{ccccccc}
\tabletypesize{\scriptsize}
\tablewidth{0pt}
\tablecaption{Positions of sources identified in at least two epochs or with a single $>5\sigma$ detection. The uncertainties for both $\alpha$ and $\delta$ are given in the units of mas. \label{tab:pos}}
\tablehead{
\colhead{VLBA} & \colhead{YSO?} & \colhead{R.A.} &\colhead{Dec.} & \colhead{Date} & \colhead{Flux} & \colhead{Field rms} \\
\colhead{} & \colhead{} & \colhead{(J2000, mas)} &\colhead{(J2000, mas)} & \colhead{(JD)} & \colhead{(mJy)} & \colhead{(mJy)}
}
\startdata
      1 & n & 5:35:29.221269$\pm$  0.361 & -5:05:44.137403$\pm$  0.491 & 2456741 &   0.593$\pm$  0.171 &   0.074 \\
      1 & n & 5:35:29.221299$\pm$  0.157 & -5:05:44.136384$\pm$  0.402 & 2456922 &   0.570$\pm$  0.174 &   0.072 \\
      1 & n & 5:35:29.221302$\pm$  0.172 & -5:05:44.136521$\pm$  0.361 & 2457080 &   0.673$\pm$  0.192 &   0.072 \\
      1 & n & 5:35:29.221309$\pm$  0.945 & -5:05:44.136624$\pm$  0.418 & 2457266 &   1.695$\pm$  0.488 &   0.079 \\
      1 & n & 5:35:29.221302$\pm$  0.254 & -5:05:44.136877$\pm$  0.485 & 2457444 &   0.702$\pm$  0.192 &   0.052 \\
 \hline
      2 & n & 5:35:54.459037$\pm$  0.021 & -5:24:37.112666$\pm$  0.038 & 2456741 &   4.856$\pm$  0.096 &   0.043 \\
      2 & n & 5:35:54.459059$\pm$  0.018 & -5:24:37.112364$\pm$  0.031 & 2456922 &   4.718$\pm$  0.112 &   0.039 \\
      2 & n & 5:35:54.459086$\pm$  0.081 & -5:24:37.113157$\pm$  0.119 & 2457083 &   5.441$\pm$  0.466 &   0.153 \\
\enddata
\end{deluxetable*}

%% file: par.tex
\begin{deluxetable*}{ccccccccc}
\tabletypesize{\scriptsize}
\tablewidth{0pt}
\tablecaption{Distance and proper motion solutions\label{tab:par}}
\tablehead{
\colhead{VLBA} & \colhead{Common} & \colhead{Spectral} & \colhead{Region} & \colhead{Parallax} & \colhead{Distance} &\colhead{$\mu_\alpha$} & \colhead{$\mu_\delta$} \\
\colhead{} & \colhead{Names} & \colhead{Type} & \colhead{} & \colhead{mas} & \colhead{pc} &\colhead{mas year$^{-1}$} & \colhead{mas year$^{-1}$}
}
\startdata
5 		&GMR A&& Trapezium & 2.673$\pm$0.112 & 374$\pm$15 &  1.81$\pm$ 0.11 & -1.62$\pm$ 0.13 \\
6\tablenotemark{b} 		&GMR F&K1IV (H97)& Trapezium & 2.567$\pm$0.051 & 389$\pm$ 7 &  2.38$\pm$ 0.08 &  0.55$\pm$ 0.14 \\
7\tablenotemark{a}		&GMR H&& Trapezium & 2.684$\pm$0.109 & 372$\pm$15 &  2.22$\pm$ 0.18 & -3.80$\pm$ 0.55 \\
8 		&GMR G&K4III (H97)& Trapezium & 2.646$\pm$0.041 & 377$\pm$ 5 &  3.82$\pm$ 0.10 &  1.60$\pm$ 0.17 \\
9\tablenotemark{b} 		&$\theta^1$ Ori E&G2IV (M12)& Trapezium & 2.557$\pm$0.051 & 391$\pm$ 7 &  1.45$\pm$ 0.03 &  1.02$\pm$ 0.08 \\
11\tablenotemark{b} 		&$\theta^1$ Ori A&B0V (H97)& Trapezium & 2.626$\pm$0.100 & 380$\pm$14 &  4.81$\pm$ 0.10 & -2.53$\pm$ 0.12 \\\hline
4/107\tablenotemark{ac}	&Brun 656&G2III (H97)& ONC & 2.708$\pm$0.210 & 369$\pm$10 &  2.36$\pm$0.69 & 0.06$\pm$1.05 \\
14 		&V1699 Ori&& ONC & 2.493$\pm$0.049 & 401$\pm$ 7 &  1.76$\pm$ 0.05 & -0.89$\pm$ 0.16 \\
\enddata
\tablenotetext{a}{Detected only in 3 epochs. Uncertainties could be underestimated.}
\tablenotetext{b}{Spectroscopic binary}
\tablenotetext{c}{Astrometric binary}
\tablenotetext{d}{Probable binary}
\tablenotetext{e}{H97=\citet{1997hillenbrand}, M12=\citet{2012Morales}, S09=\citet{2009Strassmeier}, M88=\citet{1988Marschall}, H99=\citet{1999Houk}, H13=\citet{2013Hsu}, A00=\citet{2000Alcala}, S75=\citet{1975strom}}
\end{deluxetable*}

\begin{deluxetable*}{lrclrcl}
\tabletypesize{\scriptsize}
\tablewidth{0pt}
\tablecaption{Orbital solutions for astrometric binaries\label{tab:bin}}
\tablehead{
\colhead{Parameters} & \multicolumn{3}{c}{VLBA 61/62} & \multicolumn{3}{c}{VLBA 4/107}}
\startdata
$P$ (year) & 9.50&$\pm$&0.67			&	6.27&$\pm$&0.54 \\
$T_p$ (HJD)& 2456455&$\pm$&60		&	2457355&$\pm$&200 \\
$e$ &  0.50&$\pm$&0.12				&	0.40&$\pm$&0.02 \\
$\omega$ (deg) & 92&$\pm$&20			&	268&$\pm$&7 \\
$q = M_2/M_1$ & 0.50&$\pm$&0.26		&	0.99&$\pm$&0.24 \\
$a_1$ (AU) & 2.10&$\pm$&0.51			&	(2.63&$\pm$&0.17)$/\cos i$ \\
$a_2$ (AU) & 4.19&$\pm$&0.70			&	(2.53&$\pm$&0.42)$/\cos i$ \\
$M_1$ (M$_\odot$) & 1.85&$\pm$&0.58	&	(1.70&$\pm$&0.16)$/\cos^3 i$ \\
$M_2$ (M$_\odot$) & 0.95&$\pm$&0.22	&	(1.62&$\pm$&0.38)$/\cos^3 i$ \\
$i$ (deg) & 141&$\pm$&6				&	\multicolumn{3}{c}{---} \\
$\Omega$ (deg) & 122&$\pm$&30		&	\multicolumn{3}{c}{---} \\\hline
$N_{obs}$ (primary) & \multicolumn{3}{c}{5} & \multicolumn{3}{c}{3} \\
$N_{obs}$ (secondary) & \multicolumn{3}{c}{3} & \multicolumn{3}{c}{3}
\enddata
\end{deluxetable*}

\begin{deluxetable*}{lrclrclrcl}
\tabletypesize{\scriptsize}
\tablewidth{0pt}
\tablecaption{Orbital solutions for spectroscopic binaries\label{tab:spbin}}
\tablehead{
\colhead{Parameters} & \multicolumn{3}{c}{VLBA 11} & \multicolumn{3}{c}{VLBA 19} & \multicolumn{3}{c}{VLBA 34}}
\startdata
$P$ (days) &		65.4314491&$\pm$&0.0028542 & 	33.73&$\pm$&0.030 & 		18.6561&$\pm$&0.0002  \\
$T_p$ (HJD)& 	2444195.5773&$\pm$&0.4426 & 		2444972.95&$\pm$&1.75 & 	2446010.461&$\pm$&0.080 \\
$e$ &			0.6261&$\pm$&0.0313 & 			0.12&$\pm$&0.01 & 		0.468&$\pm$&0.014 \\
$\omega$ (deg)&	183.2838&$\pm$&5.1861 & 			131.3&$\pm$&6.7 & 		118.3&$\pm$&2.4 \\
$a_1\sin i$ (AU) & 		0.16&$\pm$&0.01 & 		0.081&$\pm$&0.002 & 		0.069&$\pm$&0.003 \\
$a_2\sin i$ (AU) & 		\multicolumn{3}{c}{---} & 0.107&$\pm$&0.003 & 		0.132&$\pm$&0.009 \\
Reference & \multicolumn{3}{c}{\citet{2000Stickland}} & \multicolumn{3}{c}{\citet{1988Marschall}} & 		\multicolumn{3}{c}{\citet{1998Bolton}} \\\hline
$i$ (deg) & 		87&$\pm$&11 & 	104&$\pm$&12 & 		52&$\pm$&23 \\
$\Omega$ (deg)&	150&$\pm$&20 & 	61&$\pm$&11 & 		313&$\pm$&26 \\
$M_1$ (M$_\odot$) & 		\multicolumn{3}{c}{---} & 		0.49&$\pm$&0.10 & 		4.09&$\pm$&4.44 \\
$M_2$ (M$_\odot$) & 		\multicolumn{3}{c}{---} & 		0.37&$\pm$&0.07 & 		2.14&$\pm$&2.22
\enddata
\end{deluxetable*}


%% file: laori.tex
\begin{deluxetable*}{ccccc}
\tabletypesize{\scriptsize}
\tablewidth{0pt}
\tablecaption{Parameters of elliptical gaussian model fit for the scattered calibators towards $\lambda$ Ori\label{tab:laori}}
\tablehead{
\colhead{Name} & \colhead{$\nu$} & \colhead{$\theta_{major}$} & \colhead{$\theta_{minor}$} & \colhead{P.A.} \\
\colhead{} & \colhead{(GHz)} & \colhead{(mas)} & \colhead{(mas)} & \colhead{(deg.)}
}
\startdata
J0536+0944 & 4.98 & 16.5$\pm$0.4 & 9.9$\pm$0.1 & 139.7$\pm$ 10.2 \\
J0536+0944 & 8.42 & 5.0$\pm$0.1 & 3.5$\pm$0.1 & 150.3$\pm$ 1.7 \\
J0532+0732 & 4.98 & 9.9$\pm$0.2 & 7.2$\pm$0.1 & 173.2$\pm$2.2 \\
J0532+0732 & 8.42 & 4.3$\pm$0.1 & 3.0$\pm$0.1 & 174.0$\pm$1.2 \\
J0544+1118 & 4.98 & 6.2$\pm$0.1 & 5.1$\pm$0.1 & 163.9$\pm$5.0 \\
J0544+1118 & 8.42 & 2.9$\pm$0.1 & 2.5$\pm$0.1 & 7.3$\pm$2.0
\enddata
\end{deluxetable*}

%% file: distance_revise2.bbl
\begin{thebibliography}{}
\expandafter\ifx\csname natexlab\endcsname\relax\def\natexlab#1{#1}\fi

\bibitem[{{Abt} {et~al.}(1991){Abt}, {Wang}, \& {Cardona}}]{1991Abt}
{Abt}, H.~A., {Wang}, R., \& {Cardona}, O. 1991, \apj, 367, 155

\bibitem[{{Alcal{\'a}} {et~al.}(2000){Alcal{\'a}}, {Covino}, {Torres},
  {Sterzik}, {Pfeiffer}, \& {Neuh{\"a}user}}]{2000Alcala}
{Alcal{\'a}}, J.~M., {Covino}, E., {Torres}, G., {et~al.} 2000, \aap, 353, 186

\bibitem[{{Alves} \& {Bouy}(2012)}]{2012alves}
{Alves}, J., \& {Bouy}, H. 2012, \aap, 547, A97

\bibitem[{{Bally}(2008)}]{2008bally}
{Bally}, J. 2008, {Overview of the Orion Complex}, ed. B.~{Reipurth}, 459

\bibitem[{{Bally} {et~al.}(1987){Bally}, {Langer}, {Stark}, \&
  {Wilson}}]{1987bally}
{Bally}, J., {Langer}, W.~D., {Stark}, A.~A., \& {Wilson}, R.~W. 1987, \apjl,
  312, L45

\bibitem[{{Bolton} {et~al.}(1998){Bolton}, {Harmanec}, {Lyons}, {Odell}, \&
  {Pyper}}]{1998Bolton}
{Bolton}, C.~T., {Harmanec}, P., {Lyons}, R.~W., {Odell}, A.~P., \& {Pyper},
  D.~M. 1998, \aap, 337, 183

\bibitem[{{Bouy} {et~al.}(2014){Bouy}, {Alves}, {Bertin}, {Sarro}, \&
  {Barrado}}]{2014bouy}
{Bouy}, H., {Alves}, J., {Bertin}, E., {Sarro}, L.~M., \& {Barrado}, D. 2014,
  \aap, 564, A29

\bibitem[{{Bower} \& {Backer}(1998)}]{1998Bower}
{Bower}, G.~C., \& {Backer}, D.~C. 1998, \apjl, 496, L97

\bibitem[{{Close} {et~al.}(2013){Close}, {Males}, {Morzinski}, {Kopon},
  {Follette}, {Rodigas}, {Hinz}, {Wu}, {Puglisi}, {Esposito}, {Riccardi},
  {Pinna}, {Xompero}, {Briguglio}, {Uomoto}, \& {Hare}}]{2013Close}
{Close}, L.~M., {Males}, J.~R., {Morzinski}, K., {et~al.} 2013, \apj, 774, 94

\bibitem[{{Costero} {et~al.}(2008){Costero}, {Allen}, {Echevarr{\'{\i}}a},
  {Georgiev}, {Poveda}, \& {Richer}}]{2008Costero}
{Costero}, R., {Allen}, C., {Echevarr{\'{\i}}a}, J., {et~al.} 2008, in Revista
  Mexicana de Astronomia y Astrofisica Conference Series, Vol.~34, Revista
  Mexicana de Astronomia y Astrofisica Conference Series, 102--105

\bibitem[{{Da Rio} {et~al.}(2010){Da Rio}, {Robberto}, {Soderblom}, {Panagia},
  {Hillenbrand}, {Palla}, \& {Stassun}}]{2010dario}
{Da Rio}, N., {Robberto}, M., {Soderblom}, D.~R., {et~al.} 2010, \apj, 722,
  1092

\bibitem[{{Da Rio} {et~al.}(2016){Da Rio}, {Tan}, {Covey}, {Cottaar}, {Foster},
  {Cullen}, {Tobin}, {Kim}, {Meyer}, {Nidever}, {Stassun}, {Chojnowski},
  {Flaherty}, {Majewski}, {Skrutskie}, {Zasowski}, \& {Pan}}]{2016dario}
{Da Rio}, N., {Tan}, J.~C., {Covey}, K.~R., {et~al.} 2016, \apj, 818, 59

\bibitem[{{de Bruijne} {et~al.}(2014){de Bruijne}, {Rygl}, \&
  {Antoja}}]{2014deBruijne}
{de Bruijne}, J.~H.~J., {Rygl}, K.~L.~J., \& {Antoja}, T. 2014, in EAS
  Publications Series, Vol.~67, EAS Publications Series, 23--29

\bibitem[{{de Zeeuw} {et~al.}(2001){de Zeeuw}, {Hoogerwerf}, \& {de
  Bruijne}}]{2001deZeeuw}
{de Zeeuw}, T., {Hoogerwerf}, R., \& {de Bruijne}, J. 2001, in Astronomical
  Society of the Pacific Conference Series, Vol. 228, Dynamics of Star Clusters
  and the Milky Way, ed. S.~{Deiters}, B.~{Fuchs}, A.~{Just}, R.~{Spurzem}, \&
  R.~{Wielen}, 201

\bibitem[{{Desai} \& {Fey}(2001)}]{2001Desai}
{Desai}, K.~M., \& {Fey}, A.~L. 2001, \apjs, 133, 395

\bibitem[{{Dolan} \& {Mathieu}(2002)}]{2002Dolan}
{Dolan}, C.~J., \& {Mathieu}, R.~D. 2002, \aj, 123, 387

\bibitem[{{Duch{\^e}ne} \& {Kraus}(2013)}]{2013duchene}
{Duch{\^e}ne}, G., \& {Kraus}, A. 2013, \araa, 51, 269

\bibitem[{{Duquennoy} \& {Mayor}(1991)}]{1991Duquennoy}
{Duquennoy}, A., \& {Mayor}, M. 1991, \aap, 248, 485

\bibitem[{{Fischer} \& {Marcy}(1992)}]{1992Fischer}
{Fischer}, D.~A., \& {Marcy}, G.~W. 1992, \apj, 396, 178

\bibitem[{{Genzel} {et~al.}(1981){Genzel}, {Reid}, {Moran}, \&
  {Downes}}]{1981Genzel}
{Genzel}, R., {Reid}, M.~J., {Moran}, J.~M., \& {Downes}, D. 1981, \apj, 244,
  884

\bibitem[{{Goddi} {et~al.}(2011){Goddi}, {Humphreys}, {Greenhill}, {Chandler},
  \& {Matthews}}]{2011Goddi}
{Goddi}, C., {Humphreys}, E.~M.~L., {Greenhill}, L.~J., {Chandler}, C.~J., \&
  {Matthews}, L.~D. 2011, \apj, 728, 15

\bibitem[{{G{\'o}mez} {et~al.}(2008){G{\'o}mez}, {Rodr{\'{\i}}guez}, {Loinard},
  {Lizano}, {Allen}, {Poveda}, \& {Menten}}]{2008gomez}
{G{\'o}mez}, L., {Rodr{\'{\i}}guez}, L.~F., {Loinard}, L., {et~al.} 2008, \apj,
  685, 333

\bibitem[{{Greisen}(2003)}]{aips}
{Greisen}, E.~W. 2003, Information Handling in Astronomy - Historical Vistas,
  285, 109

\bibitem[{{Grellmann} {et~al.}(2013){Grellmann}, {Preibisch}, {Ratzka},
  {Kraus}, {Helminiak}, \& {Zinnecker}}]{2013Grellmann}
{Grellmann}, R., {Preibisch}, T., {Ratzka}, T., {et~al.} 2013, \aap, 550, A82

\bibitem[{{Gualandris} {et~al.}(2004){Gualandris}, {Portegies Zwart}, \&
  {Eggleton}}]{2004Gualandris}
{Gualandris}, A., {Portegies Zwart}, S., \& {Eggleton}, P.~P. 2004, \mnras,
  350, 615

\bibitem[{{Gudehus}(2001)}]{2001Gudehus}
{Gudehus}, D.~H. 2001, in Bulletin of the American Astronomical Society,
  Vol.~33, American Astronomical Society Meeting Abstracts \#198, 850

\bibitem[{{Gutermuth} {et~al.}(2011){Gutermuth}, {Pipher}, {Megeath}, {Myers},
  {Allen}, \& {Allen}}]{2011gutermuth}
{Gutermuth}, R.~A., {Pipher}, J.~L., {Megeath}, S.~T., {et~al.} 2011, \apj,
  739, 84

\bibitem[{{Hartmann}(2001)}]{2001hartmann}
{Hartmann}, L. 2001, \aj, 121, 1030

\bibitem[{{Hillenbrand}(1997)}]{1997hillenbrand}
{Hillenbrand}, L.~A. 1997, \aj, 113, 1733

\bibitem[{{Hillenbrand} \& {Hartmann}(1998)}]{1998Hillenbrand}
{Hillenbrand}, L.~A., \& {Hartmann}, L.~W. 1998, \apj, 492, 540

\bibitem[{{Hirota} {et~al.}(2007){Hirota}, {Bushimata}, {Choi}, {Honma},
  {Imai}, {Iwadate}, {Jike}, {Kameno}, {Kameya}, {Kamohara}, {Kan-Ya},
  {Kawaguchi}, {Kijima}, {Kim}, {Kobayashi}, {Kuji}, {Kurayama}, {Manabe},
  {Maruyama}, {Matsui}, {Matsumoto}, {Miyaji}, {Nagayama}, {Nakagawa},
  {Nakamura}, {Oh}, {Omodaka}, {Oyama}, {Sakai}, {Sasao}, {Sato}, {Sato},
  {Shibata}, {Shintani}, {Tamura}, {Tsushima}, \& {Yamashita}}]{2007hirota}
{Hirota}, T., {Bushimata}, T., {Choi}, Y.~K., {et~al.} 2007, \pasj, 59, 897

\bibitem[{{Houk} \& {Swift}(1999)}]{1999Houk}
{Houk}, N., \& {Swift}, C. 1999, in Michigan Spectral Survey, Ann Arbor, Dep.
  Astron., Univ. Michigan, Vol. 5, p. 0 (1999), Vol.~5, 0

\bibitem[{{Hsu} {et~al.}(2013){Hsu}, {Hartmann}, {Allen}, {Hern{\'a}ndez},
  {Megeath}, {Tobin}, \& {Ingleby}}]{2013Hsu}
{Hsu}, W.-H., {Hartmann}, L., {Allen}, L., {et~al.} 2013, \apj, 764, 114

\bibitem[{{Jeffries}(2007)}]{2007Jeffries}
{Jeffries}, R.~D. 2007, \mnras, 376, 1109

\bibitem[{{Kapteyn}(1918)}]{1918Kapteyn}
{Kapteyn}, J.~C. 1918, \apj, 47, 104

\bibitem[{{Kim} {et~al.}(2008){Kim}, {Hirota}, {Honma}, {Kobayashi},
  {Bushimata}, {Choi}, {Imai}, {Iwadate}, {Jike}, {Kameno}, {Kameya},
  {Kamohara}, {Kan-Ya}, {Kawaguchi}, {Kuji}, {Kurayama}, {Manabe}, {Matsui},
  {Matsumoto}, {Miyaji}, {Nagayama}, {Nakagawa}, {Oh}, {Omodaka}, {Oyama},
  {Sakai}, {Sasao}, {Sato}, {Sato}, {Shibata}, {Tamura}, \&
  {Yamashita}}]{2008kim}
{Kim}, M.~K., {Hirota}, T., {Honma}, M., {et~al.} 2008, \pasj, 60, 991

\bibitem[{{K{\"o}hler} {et~al.}(2006){K{\"o}hler}, {Petr-Gotzens},
  {McCaughrean}, {Bouvier}, {Duch{\^e}ne}, {Quirrenbach}, \&
  {Zinnecker}}]{2006kohler}
{K{\"o}hler}, R., {Petr-Gotzens}, M.~G., {McCaughrean}, M.~J., {et~al.} 2006,
  \aap, 458, 461

\bibitem[{{Kounkel} {et~al.}(2016){Kounkel}, {Hartmann}, {Tobin}, {Mateo},
  {Bailey}, \& {Spencer}}]{2016kounkela}
{Kounkel}, M., {Hartmann}, L., {Tobin}, J.~J., {et~al.} 2016, \apj, 821, 8

\bibitem[{{Kounkel} {et~al.}(2014){Kounkel}, {Hartmann}, {Loinard},
  {Mioduszewski}, {Dzib}, {Ortiz-Le{\'o}n}, {Rodr{\'{\i}}guez}, {Pech},
  {Rivera}, {Torres}, {Boden}, {Evans}, {Brice{\~n}o}, \&
  {Tobin}}]{2014kounkel}
{Kounkel}, M., {Hartmann}, L., {Loinard}, L., {et~al.} 2014, \apj, 790, 49

\bibitem[{{Kraus} {et~al.}(2009){Kraus}, {Weigelt}, {Balega}, {Docobo},
  {Hofmann}, {Preibisch}, {Schertl}, {Tamazian}, {Driebe}, {Ohnaka}, {Petrov},
  {Sch{\"o}ller}, \& {Smith}}]{2009Kraus}
{Kraus}, S., {Weigelt}, G., {Balega}, Y.~Y., {et~al.} 2009, \aap, 497, 195

\bibitem[{{Loinard} {et~al.}(2011){Loinard}, {Mioduszewski}, {Torres}, {Dzib},
  {Rodr{\'{\i}}guez}, \& {Boden}}]{gbds}
{Loinard}, L., {Mioduszewski}, A.~J., {Torres}, R.~M., {et~al.} 2011, in
  Revista Mexicana de Astronomia y Astrofisica, vol.~27, Vol.~40, Revista
  Mexicana de Astronomia y Astrofisica Conference Series, 205--210

\bibitem[{{Lombardi} {et~al.}(2011){Lombardi}, {Alves}, \&
  {Lada}}]{2011Lombardi}
{Lombardi}, M., {Alves}, J., \& {Lada}, C.~J. 2011, \aap, 535, A16

\bibitem[{{Markwardt}(2009)}]{mpfit}
{Markwardt}, C.~B. 2009, in Astronomical Society of the Pacific Conference
  Series, Vol. 411, Astronomical Data Analysis Software and Systems XVIII, ed.
  D.~A. {Bohlender}, D.~{Durand}, \& P.~{Dowler}, 251

\bibitem[{{Marschall} \& {Mathieu}(1988)}]{1988Marschall}
{Marschall}, L.~A., \& {Mathieu}, R.~D. 1988, \aj, 96, 1956

\bibitem[{{Megeath} {et~al.}(2012){Megeath}, {Gutermuth}, {Muzerolle},
  {Kryukova}, {Flaherty}, {Hora}, {Allen}, {Hartmann}, {Myers}, {Pipher},
  {Stauffer}, {Young}, \& {Fazio}}]{2012Megeath}
{Megeath}, S.~T., {Gutermuth}, R., {Muzerolle}, J., {et~al.} 2012, \aj, 144,
  192

\bibitem[{{Melis} {et~al.}(2014){Melis}, {Reid}, {Mioduszewski}, {Stauffer}, \&
  {Bower}}]{2014Melis}
{Melis}, C., {Reid}, M.~J., {Mioduszewski}, A.~J., {Stauffer}, J.~R., \&
  {Bower}, G.~C. 2014, Science, 345, 1029

\bibitem[{{Menten} {et~al.}(2007){Menten}, {Reid}, {Forbrich}, \&
  {Brunthaler}}]{2007menten}
{Menten}, K.~M., {Reid}, M.~J., {Forbrich}, J., \& {Brunthaler}, A. 2007, \aap,
  474, 515

\bibitem[{{Morales-Calder{\'o}n} {et~al.}(2012){Morales-Calder{\'o}n},
  {Stauffer}, {Stassun}, {Vrba}, {Prato}, {Hillenbrand}, {Terebey}, {Covey},
  {Rebull}, {Terndrup}, {Gutermuth}, {Song}, {Plavchan}, {Carpenter},
  {Marchis}, {Garc{\'{\i}}a}, {Margheim}, {Luhman}, {Angione}, \&
  {Irwin}}]{2012Morales}
{Morales-Calder{\'o}n}, M., {Stauffer}, J.~R., {Stassun}, K.~G., {et~al.} 2012,
  \apj, 753, 149

\bibitem[{{Muench} {et~al.}(2008){Muench}, {Getman}, {Hillenbrand}, \&
  {Preibisch}}]{2008Muench}
{Muench}, A., {Getman}, K., {Hillenbrand}, L., \& {Preibisch}, T. 2008, {Star
  Formation in the Orion Nebula I: Stellar Content}, ed. B.~{Reipurth}, 483

\bibitem[{{Neuhaeuser} {et~al.}(1998){Neuhaeuser}, {Wolk}, {Torres},
  {Preibisch}, {Stout-Batalha}, {Hatzes}, {Frink}, {Wichmann}, {Covino},
  {Alcala}, {Brandner}, {Walter}, {Sterzik}, \& {Koehler}}]{1998Neuhaeuser}
{Neuhaeuser}, R., {Wolk}, S.~J., {Torres}, G., {et~al.} 1998, \aap, 334, 873

\bibitem[{{Nishimura} {et~al.}(2015){Nishimura}, {Tokuda}, {Kimura}, {Muraoka},
  {Maezawa}, {Ogawa}, {Dobashi}, {Shimoikura}, {Mizuno}, {Fukui}, \&
  {Onishi}}]{2015nishimura}
{Nishimura}, A., {Tokuda}, K., {Kimura}, K., {et~al.} 2015, \apjs, 216, 18

\bibitem[{{Ochsendorf} {et~al.}(2015){Ochsendorf}, {Brown}, {Bally}, \&
  {Tielens}}]{2015Ochsendorf}
{Ochsendorf}, B.~B., {Brown}, A.~G.~A., {Bally}, J., \& {Tielens}, A.~G.~G.~M.
  2015, \apj, 808, 111

\bibitem[{{Ortiz-Le{\'o}n} {et~al.}(2016{\natexlab{a}}){Ortiz-Le{\'o}n},
  {Loinard}, {Kounkel}, {Dzib}, {Mioduszewski}, {Rodr{\'{\i}}guez}, {Torres},
  {Gonzalez-Lopezlira}, {Pech}, {Rivera}, {Hartmann}, {Boden}, {Evans},
  {Brice{\~n}o}, {Tobin}, \& {Galli}}]{oph2}
{Ortiz-Le{\'o}n}, G.~N., {Loinard}, L., {Kounkel}, M., {et~al.}
  2016{\natexlab{a}}, ApJ, submitted

\bibitem[{{Ortiz-Le{\'o}n} {et~al.}(2016{\natexlab{b}}){Ortiz-Le{\'o}n},
  {Loinard}, {Kounkel}, {Dzib}, {Mioduszewski}, {Rodr{\'{\i}}guez}, {Torres},
  {Pech}, {Rivera}, {Hartmann}, {Boden}, {Evans}, {Brice{\~n}o}, {Tobin}, \&
  {Galli}}]{ser}
---. 2016{\natexlab{b}}, ApJ, submitted

\bibitem[{{Pickering}(1917)}]{1917Pickering}
{Pickering}, W.~H. 1917, Harvard College Observatory Circular, 205, 1

\bibitem[{{Poveda} {et~al.}(2005){Poveda}, {Allen}, \&
  {Hern{\'a}ndez-Alc{\'a}ntara}}]{2005Poveda}
{Poveda}, A., {Allen}, C., \& {Hern{\'a}ndez-Alc{\'a}ntara}, A. 2005, \apjl,
  627, L61

\bibitem[{{Prato} {et~al.}(2002){Prato}, {Simon}, {Mazeh}, {McLean}, {Norman},
  \& {Zucker}}]{2002Prato}
{Prato}, L., {Simon}, M., {Mazeh}, T., {et~al.} 2002, \apj, 569, 863

\bibitem[{{Raghavan} {et~al.}(2010){Raghavan}, {McAlister}, {Henry}, {Latham},
  {Marcy}, {Mason}, {Gies}, {White}, \& {ten Brummelaar}}]{2010Raghavan}
{Raghavan}, D., {McAlister}, H.~A., {Henry}, T.~J., {et~al.} 2010, \apjs, 190,
  1

\bibitem[{{Reid} \& {Brunthaler}(2004)}]{2004reid}
{Reid}, M.~J., \& {Brunthaler}, A. 2004, \apj, 616, 872

\bibitem[{{Reid} \& {Honma}(2014)}]{2014Reid}
{Reid}, M.~J., \& {Honma}, M. 2014, \araa, 52, 339

\bibitem[{{Rodr{\'{\i}}guez} {et~al.}(2012){Rodr{\'{\i}}guez}, {G{\'o}mez}, \&
  {Tafoya}}]{2012rodriguez}
{Rodr{\'{\i}}guez}, L.~F., {G{\'o}mez}, Y., \& {Tafoya}, D. 2012, \mnras, 420,
  279

\bibitem[{{Sandstrom} {et~al.}(2007){Sandstrom}, {Peek}, {Bower}, {Bolatto}, \&
  {Plambeck}}]{2007sandstrom}
{Sandstrom}, K.~M., {Peek}, J.~E.~G., {Bower}, G.~C., {Bolatto}, A.~D., \&
  {Plambeck}, R.~L. 2007, \apj, 667, 1161

\bibitem[{{Schlafly} {et~al.}(2014){Schlafly}, {Green}, {Finkbeiner}, {Rix},
  {Bell}, {Burgett}, {Chambers}, {Draper}, {Hodapp}, {Kaiser}, {Magnier},
  {Martin}, {Metcalfe}, {Price}, \& {Tonry}}]{2014Schlafly}
{Schlafly}, E.~F., {Green}, G., {Finkbeiner}, D.~P., {et~al.} 2014, \apj, 786,
  29

\bibitem[{Seidelmann(1992)}]{1992Seidelmann}
Seidelmann, P.~K., ed. 1992, Explanatory Supplement to the Astronomical Almanac
  (Mill Valley, California: University Science Books)

\bibitem[{{Sherry} {et~al.}(2008){Sherry}, {Walter}, {Wolk}, \&
  {Adams}}]{2008Sherry}
{Sherry}, W.~H., {Walter}, F.~M., {Wolk}, S.~J., \& {Adams}, N.~R. 2008, \aj,
  135, 1616

\bibitem[{{Stassun} {et~al.}(2004){Stassun}, {Mathieu}, {Vaz}, {Stroud}, \&
  {Vrba}}]{2004Stassun}
{Stassun}, K.~G., {Mathieu}, R.~D., {Vaz}, L.~P.~R., {Stroud}, N., \& {Vrba},
  F.~J. 2004, \apjs, 151, 357

\bibitem[{{Stickland} \& {Lloyd}(2000)}]{2000Stickland}
{Stickland}, D.~J., \& {Lloyd}, C. 2000, The Observatory, 120, 141

\bibitem[{{Strassmeier}(2009)}]{2009Strassmeier}
{Strassmeier}, K.~G. 2009, \aapr, 17, 251

\bibitem[{{Strom} {et~al.}(1975){Strom}, {Strom}, {Carrasco}, \&
  {Vrba}}]{1975strom}
{Strom}, K.~M., {Strom}, S.~E., {Carrasco}, L., \& {Vrba}, F.~J. 1975, \apj,
  196, 489

\end{thebibliography}
